\documentclass[12pt,letterpaper]{article}

\usepackage{amsmath,amssymb,calc}
\usepackage{graphicx}
\usepackage{cite}

\newcommand\be{\begin{equation}}
\newcommand\ee{\end{equation}}
\newcommand{\bea}{\begin{eqnarray}}
\newcommand{\eea}{\end{eqnarray}}

\newcommand{\nn}{\nonumber}
\newcommand{\pd}{\partial}

\def\id{\protect{{1 \kern-.28em {\rm l}}}}

\def\id{\protect{{1 \kern-.28em {\rm l}}}}

\begin{document}

\begin{titlepage}
\begin{center}
\hfill \\
\vspace{2cm}
{\Large {\bf On Primordial Black Holes from Rapid Turns\\ in Two-field Models
\\[3mm] }}

\vskip 1.5cm
{\bf Lilia Anguelova\\
\vskip 0.5cm  {\it Institute for Nuclear Research and Nuclear Energy}\\
{\it Bulgarian Academy of Sciences, Sofia 1784, Bulgaria}\\
{\tt anguelova@inrne.bas.bg}}

\vskip 6mm

\end{center}

\vskip .1in
\vspace{1cm}

\begin{center} {\bf Abstract}\end{center}

\vspace{-1cm}

\begin{quotation}\noindent

We study rapid-turn trajectories in a class of two-field cosmological models, whose scalar manifold is the Poincar\'e disk. Background solutions in multi-field inflation, with field-space trajectories exhibiting sharp turns, can seed primordial black hole (PBH) formation. We investigate a class of exact solutions with hidden symmetry and show that they exhibit the kind of transient rapid-turn period, needed to induce PBH generation. Furthermore, we relax the symmetry condition and find, in a certain regime, modified solutions with improved behavior of the Hubble $\eta$-parameter, which preserve the desired shape of the turning rate function. Interestingly, the modified solutions describe a brief ultra-slow roll phase, followed by long-term slow roll inflation. It is notable that slow roll occurs near the center (not near the boundary) of the Poincar\'e disk, unlike in the standard $\alpha$-attractor case.

\end{quotation}

\end{titlepage}

\eject

\tableofcontents

\section{Introduction}

A period of inflationary expansion in the Early Universe is thought to be the best explanation for the large-scale homogeneity and isotropy of the present-day Universe. In standard descriptions of such an inflationary era, the accelerated expansion is due to the potential energy of one or more scalar fields, called inflatons. Although, traditionally, single-field models have garnered the most attention in the literature, multi-field inflationary models are much more natural from the perspective of fundamental descriptions of gravity. For instance, the effective actions of string compactifications usually contain many scalars and, in particular, an even number of them. There is  also a set of conjectured requirements, that an effective field theory should satisfy to be consistent with quantum gravity \cite{OOSV,OPSV}. These constraints are quite severe for single-field models \cite{GK,KVV}, but can be easily overcome in multi-field ones \cite{AP}.\footnote{It should be pointed out, though, that these swampland conjectures have been countered, for instance, in \cite{DHW,CdAMMQ,KT,AKLV,MYY}. Similarly, the recent Trans-Planckian Censorship Conjecture \cite{BV,BBLV}, relevant for cosmological inflation, has already been argued not to be necessary \cite{BdAQ}.} Last, but not least, multi-field cosmological models lead to new effects, which may be of great phenomenological interest.

A distinguishing feature of multi-field models, compared to single-field ones, is the turning rate of a trajectory in field space. This quantity characterizes the deviation of the trajectory from a geodesic. It has been understood recently \cite{CAP,BFM}, that even trajectories with rapid turns can be under perturbative control. In other words, one does not need to generalize the single-field slow-roll approximation, in the multi-field context, to slow turns as well, in order to be consistent with observations.\footnote{Actually, even in single-field models there is a particular non-slow roll regime, called constant roll, in which one can find phenomenologically viable examples; see \cite{MSY,ASW}.} Furthermore, trajectories with rapid turns can seed the generation of primordial black holes (PBH) with enough abundance to contribute appreciably to dark matter \cite{PSZ,FRPRW}.\footnote{See also \cite{KLM,DKT,PZHS,BDEEKKKNRS,CPS,ARF,BHFSSS,AAK,BF,NSS,AAK2,BKOPS,RSSS,dLDFR,dLFR,BJ,LLT,KT2,PVL,WFdLBBPR,GKS,GS,ARF2} 
for other recent studies on primordial black holes and their cosmological implications, as well as \cite{MKh,CKSY} for reviews on the topic.} In single-field models, by contrast, this is a challenge. 

It should be noted that PBH-generation in a two-field model can be successfully achieved in the context of the more traditional hybrid inflation \cite{CGB}. Nevertheless, rapid-turn models, which rely on non-geodesic motion in field space, have spurred huge interest recently due to a variety of reasons. For instance, the scalar potential in such models can be quite steep (see, for example \cite{TB}, and references therein), which is of great help in the search for suitable UV-completions. This also implies that the $\eta$-problem, plaguing supergravity realizations of inflation, can be alleviated significantly. In fact, in certain large-turn regimes it can be evaded completely \cite{CCLBNZ}. Also, it should be clear that, in view of the ever-increasing precision of cosmological observations, it is valuable to understand better different kinds of inflationary models, as well as their phenomenological implications. Our focus here will be on rapid-turn two-field models, which could realize the PBH-generating mechanism of \cite{PSZ,FRPRW}.

The arguments of \cite{PSZ,FRPRW} on the generation of primordial black holes assume that there are inflationary trajectories in field space, such that the turning rate is large in a brief period of time and (almost) vanishing outside of it. This raises the question whether it is possible to find actual solutions of the background equations of motion, which lead to such trajectories. And more broadly, in what kinds of models and under what conditions one can have trajectories, whose turning rate behaves in this manner. We begin to answer this question by studying background solutions in a certain class of two-field inflationary models with a hyperbolic scalar manifold.

Cosmological models, whose scalar manifold is hyperbolic, have been extensively studied in recent years under the name $\alpha$-attractors. The original works \cite{KLR,KLR2,KL3,CKLR} focused on the Poincar\'e disk case, while subsequent generalizations investigated models with much more involved hyperbolic surfaces as their scalar manifolds \cite{LS,BL,BL2}. In \cite{ABL}, a variety of exact solutions was found in a class of such models, by using the Noether symmetry method. This method of finding exact solutions is well-known from the study of extended theories of gravity; see for example \cite{CR,CMRS,CNP,CDeF}. It has also been applied to two-field cosmological models, for instance, in \cite{PT,PT2}. The key idea is that imposing a certain symmetry requirement constrains the scalar potential and simplifies the equations of motion.\footnote{Note that the symmetry under consideration is relevant only for the classical background solutions, not the full quantum action. So its presence is not in contradiction with any conjectures on the absence of continuous symmetries in quantum gravity.}  

Here we will focus on a class of exact solutions of \cite{ABL}, arising in the case when the scalar manifold is the Poincar\'e disk. These solutions were obtained under a ceratin simplifying ansatz for the Noether symmetry; for more general Noether symmetries in two-field models of this kind, see \cite{ABL2}. We show (combining analytical and numerical means) that the turning rate of the exact solutions under consideration behaves precisely in the desired manner. Namely, it is large during a transient period of time, while vanishing before and after it.\footnote{Note that this is a novel behavior compared to the other known in the literature (approximate) solutions in models with hyperbolic scalar manifolds, which have turning rates that are either nearly-vanishing (slow-turn inflation) or large but slowly-varying (angular inflation); see \cite{ChRS}.} This rapid-turn period induces a brief tachyonic instability of certain perturbations, triggering the kind of enhancement of the power spectrum necessary for PBH generation. Interestingly, we find that the field-space region relevant here is a small neighborhood of the center of the Poincar\'e disk, instead of its boundary as was the case in the original $\alpha$-attractors of \cite{KLR,KLR2,KL3,CKLR}. This is hugely preferable in the sense of not having to rely on super-Planckian excursions in field space.

It turns out, however, that the behavior of the Hubble $\eta$-parameter, of the exact solutions under study, is problematic phenomenologically. To remedy this, we set out to find modified solutions with a certain ansatz, inspired by the hidden symmetry of \cite{ABL}. That ansatz reduces to the hidden symmetry case, for specific parameter values, but does not respect the symmetry in general. We solve the resulting equations of motion analytically in a certain regime and show that the turning rate of the corresponding field-space trajectories behaves as before. Interestingly, we find that the modified solutions describe a brief ultra-slow roll period, followed by long-term slow roll. And, as before, the field-space region, in which this slow-roll expansion occurs, is a small neighborhood of the center of the Poincar\'e disk.

The present paper is organized as follows. In Section \ref{multi-f_dyn}, we give a brief review of the dynamics of multi-field cosmological models. Specifying to two-field models, we introduce the main characteristics of their field-space trajectories and discuss the perturbations around those trajectories. In Section \ref{2drot}, we compute the general expressions for the turning rate, and the related entropic mass of fluctuations, in models with rotationally-invariant scalar manifolds. In Section \ref{ExactSol}, we investigate a class of exact solutions with hidden symmetry, found in \cite{ABL} for the case of Poincar\'e-disk scalar manifold. We begin by outlining the derivation of these solutions, underlining some key points which will be of crucial importance in the next Section. Then, we show that their field-space trajectories have turning rate, which is large in a brief period of time and vanishing outside of it. This rapid-turn period triggers a transient tachyonic instability of the entropic perturbations. We show that, in the present case, the relevant field-space region is near the center of the Poincar\'e disk, unlike in the case of the original $\alpha$-attractors that rely on a region near the boundary. In Section \ref{NewModSol}, we find, in a certain regime, new solutions with a modified behavior of the Hubble $\eta$-parameter. We show that their turning rate function exhibits the same kind of rapid-turn phase as for the hidden symmetry solutions. As a result, again, the entropic perturbation undergoes a transient tachyonic instability. Finally, in Section \ref{PBHgen}, we discuss the implications of the modified solutions (and further corrections) for PBH generation. Appendices \ref{Rescale}, \ref{TheoremsRho} and \ref{NumEx} contain technical details and illustrations relevant for Section \ref{ExactSol}. Most importantly, in Appendix \ref{TheoremsRho} we prove that, for the hidden-symmetry solutions, the radial function of the trajectories on the Poincar\'e disk can have at most two local extrema. This restricts greatly the shape of the possible field-space trajectories.

\section{Multi-field inflationary dynamics} \label{multi-f_dyn}
\setcounter{equation}{0}

We will consider multi-field cosmological models with two scalar fields $\phi^I$ minimally coupled to gravity. The action for this system is:
\be \label{Action_gen}
S = \int d^4x \sqrt{-\det g} \left[ \frac{R}{2} - \frac{1}{2} G_{IJ} \pd_{\mu} \phi^I \pd^{\mu} \phi^J - V (\{ \phi^I \}) \right] \,\,\, ,
\ee
where $g_{\mu \nu}$ is the spacetime metric with $\mu,\nu = 0,1,2,3$ and $G_{IJ}$ is a sigma-model target space metric with $I,J = 1,2$. The usual cosmological Ansatze for the background spacetime metric and scalars are:
\be \label{metric_g}
ds^2_g = -dt^2 + a^2(t) d\vec{x}^2 \qquad , \qquad \phi^I = \phi^I_0 (t) \quad ,
\ee 
where $a(t)$ is the scale factor.

The equations of motion that (\ref{Action_gen}) implies for the background fields are the following. The scalars $\phi_0^I (t)$ satisfy:
\be \label{EoMs}
D_t \dot{\phi}^I_0 + 3 H \dot{\phi}_0^I + G^{IJ} V_J = 0 \,\,\,\,\, ,
\ee
where we have denoted $\dot{} \equiv \pd_t$, $V_J \equiv \pd_J V$, $\pd_J \equiv \pd_{\phi_0^J}$ and $H = \frac{\dot{a}}{a}$ is the Hubble parameter. Also, the derivative $D_t$ is defined by:
\be
D_t A^I \equiv \dot{\phi}_0^J \,\nabla_J A^I = \dot{A}^I + \Gamma^I_{JK} \dot{\phi}_0^J A^K \,\,\, ,
\ee
where $A^I$ is any vector in field space and $\Gamma^I_{JK}$ are the Christoffel symbols for the target space metric $G_{IJ}$. Finally, the Einstein equations can be written as:
\be \label{EinstEqs}
G_{IJ} \dot{\phi}^I_0 \dot{\phi}^J_0 = - 2 \dot{H}  \qquad  {\rm and}  \qquad  3 H^2 + \dot{H} = V \,\,\, .
\ee

\subsection{Background trajectories in two-field models}

It is useful to introduce an orthonormal basis given by the tangent and normal vectors to a background trajectory $(\phi^1_0(t),\phi^2_0(t))$ in field space, namely \cite{CAP}:
\be
T^I = \frac{\dot{\phi}^I_0}{\dot{\phi}_0}
\ee
and
\be
N_I = (\det G)^{1/2} \epsilon_{IJ} T^J \,\,\, ,
\ee
where
\be \label{phi_0_dot}
\dot{\phi}_0^2 = G_{IJ} \dot{\phi}^I_0 \dot{\phi}^J_0
\ee
and $\epsilon_{12} = - \epsilon_{21} = 1$, $\epsilon_{II} = 0$ with $I = 1,2$. Clearly, the above definitions imply that:
\be \label{Orthon}
N_I T^I = 0 \qquad , \qquad T_I T^I = 1 \qquad {\rm and} \qquad N_I N^I = 1  \quad ,
\ee
where $T_I = G_{IJ} T^J$ and $N^I = G^{IJ} N_J$.

An important quantity, whose presence is a distinguishing feature of two-field models compared to  single-field ones, is the turning rate $\Omega$ of the trajectory $(\phi^1_0(t),\phi^2_0(t))$. Its definition is \cite{AAGP}:
\be \label{Om_1}
\Omega = - N_I D_t T^I \,\,\, .
\ee
It is conceptually useful to rewrite this expression, by using (\ref{Orthon}), as:
\be \label{Om_3}
\Omega^2 = G_{IJ} (D_t T^I) (D_t T^J) = || D_t T^I ||^2 \,\,\, .
\ee
However, the simplest form of $\Omega$, for technical purposes, can be obtained by using the equations of motion. Namely, note that the projections of the field equations (\ref{EoMs}) along $T^I$ and $N_I$ give respectively:
\be
\ddot{\phi}_0 + 3 H \dot{\phi}_0 + V_{\phi} = 0 \,\,\, ,
\ee
where $V_{\phi} \equiv T^I \pd_I V$, and:
\be
N_I D_t T^I = - \frac{N_J V^J}{\dot{\phi}_0} \,\,\, .
\ee
Using the last relation in (\ref{Om_1}), one finds:
\be \label{Om_2}
\Omega = \frac{N_I V^I}{\dot{\phi}_0} \,\,\, .
\ee
This expression for the turning rate turns out to be the most useful one computationally. 

In the multi-field context, the slow-roll parameters can be defined as \cite{CAP}:
\be \label{SR_par}
\varepsilon = - \frac{\dot{H}}{H^2} \qquad , \qquad \eta^I = - \frac{1}{H \dot{\phi}_0} D_t \dot{\phi}_0^I \,\,\, .
\ee
One can expand $\eta^I$ in the above orthonormal basis:
\be
\eta^I = \eta_{\parallel} T^I + \eta_{\perp} N^I \,\,\, ,
\ee
where the coefficients are:
\be \label{eta_PP}
\eta_{\parallel} = - \frac{\ddot{\phi}_0}{H \dot{\phi}_0} \qquad {\rm and} \qquad \eta_{\perp} = \frac{N_I V^I}{\dot{\phi}_0 H} \,\,\, .
\ee
Clearly, (\ref{Om_2}) and (\ref{eta_PP}) imply:
\be
\Omega = \eta_{\perp} H \,\,\, .
\ee

Notice that the above definitions of $\varepsilon$ and $\eta_{\parallel}$ coincide with the standard slow-roll parameters for single-field inflation with inflaton $\phi_0 (t)$. As in that case, one can define the slow-roll approximation by:
\be
\varepsilon <\!\!< 1 \qquad , \qquad |\eta_{\parallel}| <\!\!<1 \,\,\, .
\ee
Note however, that the magnitude of the remaining parameter, the dimensionless turning rate $\eta_{\perp} = \Omega / H$\,, does not have to be small in order to have perturbative control of the computations \cite{CAP,BFM}; see also \cite{FGSPRPR}. This will be of crucial importance in the following.

\subsection{Adiabatic and entropic perturbations}

Important aspects of the physics of an inflationary model are determined by the dynamics of scalar perturbations around the homogeneous background. One can define the perturbations of the scalar fields via the decomposition:
\be \label{phi_decomp}
\phi^I (t, \vec{x}) = \phi^I_0 (t) + \delta \phi^I (t, \vec{x}) \,\,\, .
\ee
As for the metric perturbations, it is convenient to work in comoving gauge, in which they are contained in the spatial part $g_{ij}$, with $i,j = 1,2,3$\,, as:
\be
g_{ij} (t, \vec{x}) = a^2(t) \left[ ( 1 + 2 \zeta ) \delta_{ij} + h_{ij} \right] \,\,\, ,
\ee 
where $\zeta = \zeta (t,\vec{x})$ is the curvature perturbation and $h_{ij}$ - the tensor fluctuations.

Generically, the fluctuations in (\ref{phi_decomp}) can be decomposed as $\delta \phi^I = (\delta \phi_{\parallel}) T^I + (\delta \phi_{\perp}) N^I$, where $\delta \phi_{\parallel}$ is the adiabatic perturbation and $\delta \phi_{\perp}$ is the isocurvature (equivalently, entropic) one. However, in comoving gauge $\delta \phi_{\parallel}$ vanishes identically. Therefore, $Q_s \equiv \delta \phi_{\perp}$ and $\zeta$ are the only independent scalar degrees of freedom. The quadratic Lagrangian for these fields is \cite{GWBM,GNvT,SS,LRP}:
\be \label{L_fluct}
{\cal L}_{pert} = a^3 \left[ \,\varepsilon \!\left( \dot{\zeta}^2 - \frac{\pd_i \zeta \pd^i \zeta}{a^2} \right) + 2 \dot{\phi}_0 \eta_{\perp} \,\dot{\zeta} Q_s + \frac{1}{2} \!\left( \dot{Q}_s^2 - \frac{\pd_i Q_s \pd^i Q_s}{a^2} - m_s^2 Q_s^2 \right) \right] \, ,
\ee 
where $m_s^2$ is the effective entropic mass of the perturbations. It is given by the following expression:
\be \label{m2_eff_gen}
m_s^2 = N^I N^J V_{;IJ} - \Omega^2 + \varepsilon H^2 \mathcal{R} \,\,\, ,
\ee
where $V_{;IJ} = \pd_I \pd_J V - \Gamma^K_{IJ} V_K$ and $\mathcal{R}$ is the Ricci scalar of the sigma-model metric $G_{IJ}$. 

Recently it was argued that a brief period of instability, driven by a large turning rate, can cause enough enhancement of the power spectrum of the curvature perturbation, in order to generate a significant amount of primordial black holes (PBH) \cite{PSZ,FRPRW}. The mechanism for PBH production proposed in these works relies on a transient inflationary period of time $\Delta t$ with:
\be
\eta_{\perp}^2 >\!\!> 1 \,\,\, .
\ee
An important role is played by the interaction term $\dot{\zeta} Q_s$ in (\ref{L_fluct}), whose strength notably depends on $\eta_{\perp}$\,. In \cite{PSZ} a so called ultra-light case (see \cite{AAGP}) was studied and also it was assumed that $\eta_{\perp} = \Omega/H$ is constant during $\Delta t$ and zero otherwise. On the other hand, \cite{FRPRW} considered an ansatz such that $m_s^2$ becomes large and negative during the rapid-turn period $\Delta t$\,, while being small and positive before and after it; for convenience, they also assumed that the function $\eta_{\perp} (t)$ is described by a Gaussian. In either case, the transient tachyonic instability caused by a large $\eta_{\perp}$ induces an exponential growth of the power spectrum, which then leads to the desired PBH generation.

Both works \cite{PSZ,FRPRW} make the ad hoc assumption that there are background trajectories $\phi_0^I (t)$\,, such that the resulting turning rate, as a function of time, behaves as described above. Our goal, in the following, will be to investigate the question of whether and how one can find this kind of trajectories. More precisely, we will show that there are solutions of the background equations of motion for a certain class of two-field models, whose trajectories have a turning rate that is large in a brief period of time and vanishing before and after it. Interestingly, we will find that the functions $\Omega (t)$\,, equivalently $\eta_{\perp} (t)$\,, and $m_s^2 (t)$\,, which result from these trajectories, behave rather similarly to the corresponding assumptions of \cite{FRPRW}.

\section{2d rotationally invariant target spaces} \label{2drot}
\setcounter{equation}{0}

The class of two-field models, whose background field equations will be of interest to us, is obtained by taking the target space metric $G_{IJ}$ in (\ref{Action_gen}) to be hyperbolic and rotationally invariant. Before specifying the form of $G_{IJ}$ further, it will be useful first to derive the general expressions for the turning rate $\Omega$ and entropic mass $m_s^2$\,, that follow simply from requiring rotational invariance of the sigma-model target space.

Let us, for convenience, denote:
\be \label{Backgr_id}
\phi^1_0 (t) \equiv \varphi (t) \qquad {\rm and} \qquad \phi^2_0 (t) \equiv \theta (t)
\ee
from now on. Then $G_{IJ}$\,, as any rotationally invariant metric, can be written in the form:
\be \label{Gmetric}
ds^2_{G} = d\varphi^2 + f(\varphi) d\theta^2 \,\,\, ,
\ee
where the function $f (\varphi) \ge 0$ \,for any $\varphi$\,. The only non-vanishing Christoffel symbols $\Gamma^I_{JK}$ of the metric (\ref{Gmetric}) are:
\be \label{Chris_f}
\Gamma^{\varphi}_{\theta \theta} = - \frac{f'}{2} \qquad {\rm and} \qquad \Gamma^{\theta}_{\varphi \theta} = \frac{f'}{2 f} \quad ,
\ee
where $f' \equiv \pd_{\varphi} f$. Hence, the Ricci scalar $\mathcal{R}$ of (\ref{Gmetric}) is:
\be \label{Ricci_sc_f}
\mathcal{R} = \frac{f'^2}{2 f^2} - \frac{f''}{f} \,\,\, .
\ee
Note that, as alluded to above, we will be interested in the case when $G_{IJ}$ is the metric on a hyperbolic surface. In that case, by definition ${\cal R} = const < 0$\,. Clearly, this condition restricts the form of the function $f (\varphi)$\,. We will come back to this point in the next Section.

Now, in view of (\ref{Backgr_id})-(\ref{Gmetric}), relation (\ref{phi_0_dot}) becomes:
\be \label{phi_0}
\dot{\phi}_0^2 = \dot{\varphi}^2 + f (\varphi) \dot{\theta}^2 \,\,\, .
\ee
Using this and (\ref{Chris_f}), we find from (\ref{Om_1}) that the turning rate is:
\be \label{Om_f0}
\Omega = - \frac{\sqrt{f}}{\left( \dot{\varphi}^2 + f \dot{\theta}^2 \right)} \left[ \dot{\theta} \!\left( \ddot{\varphi} - \frac{f'}{2} \dot{\theta}^2 \right) - \dot{\varphi} \!\left( \ddot{\theta} + \frac{f'}{f} \dot{\varphi} \dot{\theta} \right) \right] \,\,\, .
\ee
One can obtain a simpler expression for $\Omega$ from (\ref{Om_2}):
\be \label{Om_f}
\Omega = \frac{\sqrt{f}}{\left( \dot{\varphi}^2 + f \dot{\theta}^2 \right)} \left[ \dot{\theta} \pd_{\varphi} V - \frac{\dot{\varphi}}{f} \pd_{\theta} V \right] \,\,\, .
\ee
It is easy to verify that (\ref{Om_f0}) and (\ref{Om_f}) are equivalent to each other for solutions of the equations of motion, as should be the case. Indeed, for that purpose, note that the field equations (\ref{EoMs}) now have the form:
\be \label{ScalarEoMs}
\ddot{\varphi} - \frac{f'}{2} \dot{\theta}^2 + 3 H \dot{\varphi} + \pd_{\varphi} V = 0 \qquad , \qquad \ddot{\theta} + \frac{f'}{f} \dot{\varphi} \dot{\theta} + 3 H \dot{\theta} + \frac{1}{f} \pd_{\theta} V = 0  \quad .
\ee
For completeness, let us also write down the Einstein equations (\ref{EinstEqs}) for the present case:
\be \label{EinstEq}
\dot{\varphi}^2 + f \dot{\theta}^2 = - 2 \dot{H} \qquad , \qquad 3H^2 + \dot{H} = V \quad .
\ee
Finally, computing the entropic mass from (\ref{m2_eff_gen}), we obtain:
\be \label{m2_eff_f_gen}
m_s^2 = N_{\varphi}^2 \pd_{\varphi}^2 V + 2\frac{N_{\varphi} N_{\theta}}{f} \left( \pd_{\varphi} \pd_{\theta} V - \frac{f'}{2f} \pd_{\theta} V \right) + \frac{N_{\theta}^2}{f^2} \left( \pd_{\theta}^2 V + \frac{f'}{2} \pd_{\varphi} V \right) - \Omega^2 - \dot{H} \mathcal{R} \,\,\, ,
\ee
where
\be
N_{\varphi} = f^{1/2} \frac{\dot{\theta}}{\dot{\phi}_0} \qquad {\rm and} \qquad N_{\theta} = - f^{1/2} \frac{\dot{\varphi}}{\dot{\phi}_0}
\ee
with $\dot{\phi}_0$ given by (\ref{phi_0}).

It is important to note that one can have genuine two-field trajectories $\left( \varphi(t) , \theta(t) \right)$, which solve the background equations of motion, even when the potential $V(\varphi,\theta)$ does not depend on one of the scalars; see \cite{ARB,MM,ABL}, for example. In that case, the expressions for the turning rate and, especially, the entropic mass simplify considerably. Namely, under the assumption: 
\be
\pd_{\theta} V = 0 \,\,\, ,
\ee
we have from (\ref{Om_f}):
\be \label{Om_PD}
\Omega = \frac{\sqrt{f}}{\left( \dot{\varphi}^2 + f \dot{\theta}^2 \right)} \,\dot{\theta} \,\pd_{\varphi} V \,\,\, ,
\ee
while (\ref{m2_eff_f_gen}) acquires the form:
\be 
\label{m_s_f}
m_s^2 = M^2_V - \Omega^2 + \varepsilon H^2 \mathcal{R} \,\,\, ,
\ee
where:
\be \label{MV}
M^2_V \equiv \frac{ f \dot{\theta}^2 \pd_{\varphi}^2 V + \frac{f'}{2 f} \dot{\varphi}^2 \pd_{\varphi} V }{ ( \dot{\varphi}^2 + f \dot{\theta}^2 ) } \,\,\, .
\ee
Note that, in the slow-roll approximation (in which $\varepsilon <\!\!< 1$), the last term in $m_s^2$ can be neglected for sigma-model target spaces with ${\cal R} = const$\,. This will provide a further simplification for us, as we will consider hyperbolic surfaces, which have constant and negative  ${\cal R}$ by definition as mentioned above.

\section{Exact solutions with hidden symmetry} \label{ExactSol}
\setcounter{equation}{0}

In this Section we will investigate a class of exact solutions to the background equations of motion (\ref{ScalarEoMs})-(\ref{EinstEq}) found in \cite{ABL}. These solutions are obtained by taking the sigma-model target space with metric $G_{IJ}$ to be the Poincar\'e disk and the scalar potential $V$ to be independent of $\theta$\,. More precisely, the function $f$ in the metric (\ref{Gmetric}) has the form:
\be \label{f_D}
f (\varphi) \,= \,\frac{8}{3} \,\sinh^2 \!\left( \sqrt{\frac{3}{8}} \,\varphi \right) \,\,\,\, ,
\ee
which in particular implies from (\ref{Ricci_sc_f}) that $\mathcal{R} = - \frac{3}{4}$\,,
and the scalar potential is given by:
\be \label{Pot}
V (\varphi , \theta) \,= \,V_0 \,\cosh^2 \!\left( \sqrt{\frac{3}{8}} \,\varphi \right) \,\,\,\, ,
\ee
where $V_0 = const$\,. With these functions $f$ and $V$, the scalar field equations of motion (\ref{ScalarEoMs}) are solved by:\footnote{For convenience, here we use slightly different notation than in \cite{ABL}. Also, without any loss of generality, we have fixed three other integration constants that were present there; see Appendix \ref{Rescale}.}
\bea \label{Sols_asc}
a (t) &=& \left[ u^2 - \left( v^2 + w^2 \right) \right]^{1/3} \,\,\,\, , \nn \\
\varphi (t) &=& \sqrt{\frac{8}{3}} \,{\rm arccoth} \!\left( \sqrt{\frac{u^2}{v^2 + w^2}} \,\,\right) \,\,\,\, , \nn \\
\theta (t) &=& {\rm arccot} \!\left( \frac{v}{w} \right) \,\,\,\, ,
\eea
where the functions $u=u(t)$, $v=v(t)$ and $w=w(t)$ are:
\bea \label{Sols_uvw}
u (t) &=& C^u_1 \sinh \!\left( \kappa \,t \right) + C^u_0 \cosh \!\left( \kappa \,t \right) \,\,\, , \,\,\, \kappa \equiv \frac{1}{2} \sqrt{3 V_0} \quad \,, \nn \\
v (t) &=& C_1^v \,t + C_0^v \qquad {\rm and} \qquad w (t) = C_1^w \,t + C_0^w \quad \,,
\eea
with $C^{u,v,w}_{0,1} = const$\,. To ensure that the Einstein equations (\ref{EinstEq}) are also satisfied, one has to impose the following constraint among the integration constants in (\ref{Sols_uvw}):
\be \label{Constr_s}
(C_1^v)^2 + (C_1^w)^2 \,= \,\kappa^2 \left[ (C_1^u)^2 - (C_0^u)^2 \right] \,\,\, .
\ee

For future purposes, it will be useful to explain how this six-parameter family of solutions was obtained. So, before we turn to investigating it, we will sketch briefly its derivation, emphasizing only the points that will be needed in the next Section; the full details can be found in \cite{ABL}. Note that outlining the derivation of the solutions (\ref{Sols_asc})-(\ref{Constr_s}) will be helpful in clarifying their structure.

\subsection{Derivation via Noether method}

Now we will explain how to obtain the solutions (\ref{Sols_asc})-(\ref{Constr_s}) by using the Noether symmetry method. Note that this is simply a technique for finding classical solutions by imposing a certain symmetry requirement.\footnote{It has long been used in the context of extended theories of gravity \cite{CR,CMRS,CNP,CDeF}.} This symmetry, however, does not extend to the full quantum system. So there is no issue regarding any of the arguments or conjectures about the nonexistence of continuous symmetries in quantum gravity.

As a first step, let us substitute the homogeneous background ansatze (\ref{metric_g}), together with the sigma-model metric (\ref{Gmetric}), into the action (\ref{Action_gen}). After an integration by parts, one finds the following Lagrangian density per unit spatial volume:
\be \label{L_class_mech}
{\cal L} \,= \,- 3 a \dot{a}^2 + \frac{a^3 \dot{\varphi}^2}{2} + \frac{a^3 f (\varphi) \,\dot{\theta}^2}{2} - a^3 V(\varphi,\theta) \,\,\, .
\ee
One can view this as a classical mechanics Lagrangian for the set of generalized coordinates $\{Q_i\} \equiv \{a,\varphi,\theta\}$\,. Then, one can require that (\ref{L_class_mech}) has a Noether symmetry. Namely, one can impose the condition $L_X {\cal L} = 0$\,, where $L_X$ is the Lie derivative along the vector field $X = \lambda_{Q_i} \pd_{Q_i} + \dot{\lambda}_{Q_i} \pd_{\dot{Q}_i}$ with $\lambda_{Q_i}$ being functions of the generalized coordinates $a,\varphi,\theta$. The condition $L_X {\cal L} = 0$ actually decomposes into a set of coupled partial differential equations. For a given function $f(\varphi)$\,, this system of PDEs determines not only the functions $\lambda_{a,\varphi,\theta} (a,\varphi,\theta)$ but also the potential $V(\varphi,\theta)$; for more details, see \cite{ABL}. In other words, requiring the existence of a symmetry restricts the form of the potential. Once we have solved the condition $L_X {\cal L} = 0$\,, we can perform a coordinate transformation $(a,\varphi,\theta) \rightarrow (u,v,w)$ in (\ref{L_class_mech}) to new generalized coordinates that are adapted to the symmetry. The equations of motion, which result from the transformed Lagrangian, are considerably simplified and that enables one to find many exact solutions. 

To be able to write down explicitly the coordinate transformation relevant for us, we need to specify the function $f$. So let us now turn to the choice of a target space metric $G_{IJ}$\,. As already mentioned, we will take the target space to be the Poincar\'e disk. This is the simplest example of a hyperbolic surface. It should be noted that inflationary models with hyperbolic target spaces, called $\alpha$-attractors, are very appealing for phenomenological reasons. The original works \cite{KLR,KLR2,KL3,CKLR} focused on the Poincar\'e disk as the target space, while subsequent investigations generalized these models to the case of arbitrary hyperbolic surfaces \cite{LS,BL,BL2}. In the standard conventions of the original $\alpha$-attractor literature, the Poincar\'e disk metric has the form:
\be \label{Gmetric_al}
ds^2_D \,= \,6\alpha \,\frac{dz d\bar{z}}{(1-z\bar{z})^2} \,\,\, ,
\ee
where $z=\rho e^{i \theta}$ with $\rho \in [0,1)$ being the canonical radial variable on the disk and $\alpha$ - an arbitrary parameter. The metric (\ref{Gmetric_al}) can be mapped into the form (\ref{Gmetric}) by the change of variable:
\be
\rho \,= \,\tanh \!\left( \frac{\varphi}{\sqrt{6 \alpha}} \right) \,\,\, .
\ee
The resulting form of the function $f$ is:
\be \label{f_al}
f (\varphi) \,= \,\frac{3 \alpha}{2} \,\sinh^2 \!\left( \sqrt{\frac{2}{3 \alpha}}\,\varphi \right) \,\,\, .
\ee

With this choice of $f (\varphi)$\,, it was shown in \cite{ABL} that the existence of a hidden symmetry requires $\alpha = \frac{16}{9}$\,, resulting in the expression (\ref{f_D}). Then, \cite{ABL} found the following symmetry-compatible potential:\footnote{In (\ref{Pot_full})-(\ref{uvw_coord_tr}) we have substituted (\ref{C1C2Cw}) inside the results of \cite{ABL} and, also, we have taken into account the slight change of notation, compared to \cite{ABL}, that we introduced in (\ref{Sols_asc})-(\ref{Sols_uvw}).}
\be \label{Pot_full}
V (\varphi,\theta) \,= \,V_0 \,\cosh^2 \!\left( \sqrt{\frac{3}{8}}\,\varphi \right) \coth^{-n} \!\left( \sqrt{\frac{3}{8}}\,\varphi \right) \cos^n \!\theta \,\,\, ,
\ee
where $n$ is an arbitrary number, as well as the symmetry-adapted generalized coordinates:
\bea \label{uvw_coord_tr}
u &=& a^{3/2} \cosh \!\left( \sqrt{\frac{3}{8}}\,\varphi \right) \nn \\
v &=& a^{3/2} \sinh \!\left( \sqrt{\frac{3}{8}}\,\varphi \right) \cos \theta \nn \\
w &=& a^{3/2} \sinh \!\left( \sqrt{\frac{3}{8}}\,\varphi \right) \sin \theta \,\,\, .
\eea
Clearly, (\ref{Sols_asc}) is the inverse of (\ref{uvw_coord_tr}). In view of the considerations of Section \ref{2drot}, from now on we will take $n=0$\,, in order to keep the computations more manageable. Then, the potential (\ref{Pot_full}) reduces exactly to the expression in (\ref{Pot}).

Substituting (\ref{f_D}) and (\ref{Pot}) inside (\ref{L_class_mech}) and changing variables to $(u,v,w)$, the Lagrangian acquires the form:
\be \label{Lagr_s}
{\cal L} \,= \,- \frac{4}{3} \dot{u}^2 + \frac{4}{3} \dot{v}^2 + \frac{4}{3} \dot{w}^2 - V_0 u^2 \,\,\, .
\ee
The solutions of the Euler-Lagrange equations of (\ref{Lagr_s}) are given precisely by (\ref{Sols_uvw}). Substituting the latter in (\ref{Sols_asc}), one obtains solutions of the scalar field equations (\ref{ScalarEoMs}). To ensure that the Einstein equations (\ref{EinstEq}) are satisfied as well, one has to impose, in addition, the Hamiltonian constraint:
\be
E_{\cal L} = 0 \qquad , \qquad E_{\cal L} \equiv \frac{\pd {\cal L}}{\pd \dot{u}} \dot{u} + \frac{\pd {\cal L}}{\pd \dot{v}} \dot{v} + \frac{\pd {\cal L}}{\pd \dot{w}} \dot{w} - {\cal L} \quad .
\ee
This condition is well-known to result only in a relation between the integration constants, which in the present case is the constraint (\ref{Constr_s}). For the full derivation, as well as for many more exact solutions of the background equations of motion (\ref{ScalarEoMs})-(\ref{EinstEq}), see \cite{ABL}.

\subsection{Turning rate and entropic mass} \label{TREM}

Now we will investigate the turning rate $\Omega (t)$ and entropic mass $m_s^2 (t)$ for the trajectories (\ref{Sols_asc})-(\ref{Constr_s}). Recall that these trajectories solve the field equations (\ref{ScalarEoMs})-(\ref{EinstEq}), when the functions $f (\varphi)$ and $V (\varphi)$ are given by (\ref{f_D}) and (\ref{Pot}), respectively. Substituting (\ref{f_D})-(\ref{Sols_asc}) in (\ref{Om_PD}) and (\ref{MV}), we find:
\be \label{Om_uvw}
\Omega \,= \frac{3V_0}{4} \,\frac{u \,(v \dot{w} - \dot{v} w) \,\sqrt{u^2-w^2-v^2}}{\left[ (v \dot{u} - \dot{v} u)^2 + (w \dot{u} - \dot{w} u)^2 - (v \dot{w}-\dot{v} w)^2 \right]}
\ee
and
\be \label{m2_uvw}
M_V^2 = \frac{3V_0}{4} \,\frac{ \left\{ u^2 \left[ (v \dot{u} - \dot{v} u)^2 + (w \dot{u} - \dot{w} u)^2 \right] - (v^2 + w^2) (v \dot{w}-\dot{v} w)^2 \right\}}{(u^2 - v^2 - w^2) \left[ (v \dot{u} - \dot{v} u)^2 + (w \dot{u} - \dot{w} u)^2 - (v \dot{w}-\dot{v} w)^2 \right]} \,\,\, ,
\ee
respectively. Unfortunately, using (\ref{Sols_uvw}) in the above expressions leads to rather messy and unmanageable functions. So we will aid our investigation by numerical considerations, in order to elucidate the behavior of the expressions in (\ref{Om_uvw}) and (\ref{m2_uvw}).

To understand the types of trajectories, that the solutions under consideration can have, it will be useful to work with the canonical radial coordinate $\rho$ on the Poincar\'e disk, whose range is $\rho \in [0,1)$\,. As recalled above, its relation to the field $\varphi$ is given by \cite{ABL}:
\be \label{rho_phi}
\rho = \tanh \!\left( \frac{1}{8} \sqrt{6} \,\varphi \right) \,\, .
\ee
Substituting (\ref{Sols_asc}), this implies:
\be \label{rho_uvw}
\rho (t) = \tanh \!\left( \frac{1}{2} \,\sqrt{\frac{3}{8}} \,\varphi \right) = \frac{\sqrt{v^2 + w^2}}{\sqrt{u^2 - v^2 - w^2} + \sqrt{u^2}} \,\,\,\, ,
\ee
where we have used the relation \, $\tanh \left( \frac{1}{2} \,{\rm arccoth} \beta \right) = \frac{1}{\sqrt{\beta^2 - 1} \pm \sqrt{\beta^2}}$ \, valid for any $\beta$\,.\footnote{One can prove this relation by using \,$\sinh(2 \alpha) = \frac{2 \coth \alpha}{\coth^2 \alpha - 1}$ \,, which is valid for any $\alpha$.}
From (\ref{rho_uvw}) one can see that the condition for a local extremum of the function $\rho (t)$ is equivalent with:
\be \label{rho_extr_eq}
( v^2 + w^2 ) \,\dot{u} - \left( v \dot{v} + w \dot{w} \right) u = 0 \,\,\, .
\ee

In Appendix \ref{TheoremsRho} we have proven that equation (\ref{rho_extr_eq}), with $u(t)$, $v(t)$ and $w(t)$ given by (\ref{Sols_uvw}), can have at most two roots. In other words, there are only three types of trajectories: 1) with two local extrema, 2) with one local extremum or 3) with no local extrema of $\rho (t)$. Furthermore, for the trajectories of type 1), the earlier (in time) local extremum is a minimum, while the later one is a local maximum of $\rho (t)$. And also, for the trajectories of type 2), the single local extremum is a maximum of $\rho (t)$\,. We have illustrated all three kinds of trajectories on Figures \ref{RTh_Cv_0_Tps} and \ref{Cv_nRho} in Appendix \ref{NumEx}. From the plots on Figures \ref{Cv_nRho} and \ref{Cv_n_Om_Ms} there, it is easy to understand that trajectories of type 3), i.e. with no local extrema of $\rho (t)$\,, lead to a significantly lower and wider peak of the turning rate (\ref{Om_uvw}), than trajectories of types 1) and 2).\footnote{Note that the overall sign of $\Omega (t)$ is fixed for each trajectory (as explained in Appendix \ref{NumEx}). The angular motion along trajectories with $\Omega (t) > 0$ is in an anti-clockwise direction, with $\Omega (t)$ having a maximum, whereas the angular motion along trajectories with $\Omega (t) < 0$ is in a clockwise direction, with $\Omega (t)$ having a minimum. To encompass both situations simultaneously, here we will refer to the point of greatest $|\Omega|$ as the ``peak", instead of the maximum or minimum.} And also, the greater the peak of $\Omega (t)$ is, the better its position, $t_{peak}$\,, is approximated by the position, $t_{max}$, of the local maximum of $\rho (t)$\,, as observed in Appendix \ref{NumEx}. We have verified these conclusions numerically on many examples, although, in hindsight, they should not be surprising. Indeed, trajectories with no local $\rho (t)$-extrema have no sharp turns, whereas, naturally, the highest and sharpest peaks of $|\Omega (t)|$ are achieved for the trajectories with the sharpest turns. This correlation, between the form of $\Omega (t)$ and the shape of the corresponding trajectory, is clearly visible on many Figures in Appendix \ref{NumEx}. Note also that, in all examples, the entropic mass $m_s^2 (t)$\,, obtained from substituting (\ref{Om_uvw})-(\ref{m2_uvw}) in (\ref{m_s_f}), develops the desired brief tachyonic instability, as illustrated on multiple Figures in Appendix \ref{NumEx}. 

From Appendices \ref{TheoremsRho} and \ref{NumEx}, one can realize that, although there are many examples with $t_{peak} > 1$ (as illustrated on Figures \ref{Tp_5} and \ref{Tp_10} there), by choosing randomly the values of the constants in (\ref{Sols_uvw})-(\ref{Constr_s}), especially with $\kappa > 1$\,, one has a good chance of obtaining $t_{peak} < 1$\,. Indeed, in Appendix \ref{TheoremsRho} we proved that $t_{max} < t_+^s$\,, while in Appendix \ref{NumEx} we saw that $t_{peak} \sim t_{max}$ with the approximation $t_{peak} \approx t_{max}$ being better (and, in fact, very accurate) for higher peaks. Therefore, from the $t_+^s$ expression in (\ref{t_s}) it is clear that finding examples with $t_{peak} > 1$ requires some care, although there are infinitely many of them. Searching for such examples is greatly facilitated by making the following observation. In all cases on Figures \ref{Tp_5} and \ref{Tp_10} (as well as any other numerical example with $t_{peak} > 1$\,, that we have explored), the following inequalities are satisfied in a large neighborhood of $t = t_{peak}$\,:
\be \label{Approx_reg_0}
|u|,|\dot{u}| \,>\!\!> \,|v|, |w|, |\dot{v}|, |\dot{w}|
\ee 
In fact, in all those cases, the regime (\ref{Approx_reg_0}) is reached numerically rather fast, i.e. at some moment $t_0 < 1$\,. Furthermore, the greater $t_{peak}$ is, the better-satisfied (\ref{Approx_reg_0}) is throughout the entire trajectory, except for a brief period of time at the beginning (i.e., close to $t=0$). Note also that the examples with larger $t_{peak}$ are better phenomenologically regarding the slow roll approximation. Namely, although the function $\varepsilon (t)$ generally satisfies the slow roll condition \,$\varepsilon <\!\!< 1$\, around $t \approx t_{peak}$\,, it is larger for the examples with smaller $t_{peak}$\,. More than that, for the cases with $t_{peak} < 1$\,, the $\varepsilon$ parameter can be quite large at small $t$ (even greater than $1$), decreasing enough only rather close to $t \sim t_{peak}$\,. So, for phenomenological reasons, from now on we will concentrate on the cases with $t_{peak} > 1$\,. Note that in those cases, in addition to (\ref{Approx_reg_0}), one also has numerically:
\be \label{Ineq_uvw_approx}
(v \dot{u} - \dot{v} u)^2 + (w \dot{u} - \dot{w} u)^2 \, >\!\!> \, (v \dot{w} - \dot{v} w)^2 
\ee
throughout the entire trajectory, except for a very small neighborhood of $t=0$\,. This inequality is consistent with, but not necessarily a consequence of, the inequalities in (\ref{Approx_reg_0}).\footnote{In the numerical examples with $t_{peak} < 1$\,, neither (\ref{Approx_reg_0}) nor (\ref{Ineq_uvw_approx}) is satisfied.} In the following we will refer to (\ref{Approx_reg_0})-(\ref{Ineq_uvw_approx}) as the large-$u$ limit.

In addition to being phenomenologically preferable, the large-$u$ limit provides a corner of parameter space, in which it is possible to obtain analytical estimates for the main characteristics of the function $\Omega (t)$. Indeed, within the approximations (\ref{Approx_reg_0})-(\ref{Ineq_uvw_approx}), the expression (\ref{Om_uvw}) simplifies to:
\be \label{Om_lu_h_sym}
\Omega \approx \frac{3V_0}{4} \,\frac{u^2 \,(v \dot{w} - \dot{v} w)}{\left[ (v \dot{u} - \dot{v} u)^2 + (w \dot{u} - \dot{w} u)^2 \right]} \,\, .
\ee
Now, since we are looking for examples with (relatively) large $t_{peak}$\,, we will assume that the negative exponent inside $u(t)$ in (\ref{Sols_uvw}) is negligible for the present purposes.\footnote{This can be ensured by taking $\kappa > 1$ and/or $C_1^u\approx C_0^u$ (although $C_1^u\neq C_0^u$). In fact, both of these conditions are satisfied in the examples illustrated on Figures \ref{Tp_5} and \ref{Tp_10} in Appendix \ref{NumEx}, where one has $|C_1^u - C_0^u| < 10^{-2}$\,. Note that, for the solutions (\ref{Sols_asc})-(\ref{Sols_uvw}), taking $\kappa > 1$ allows for more than one e-fold of expansion to occur by the time $t=t_{peak}$\,.} With this assumption, one finds that (\ref{Om_lu_h_sym}) has a single extremum at:
\be \label{t_peak_l}
t_{peak} \, \approx \, \frac{1}{\kappa} - \frac{C_1^vC_0^v+C_1^wC_0^w}{(C_1^v)^2+(C_1^w)^2} \,\,\, .
\ee
Substituting this in (\ref{Om_lu_h_sym}), we obtain:
\be \label{Om_hid_sym_p}
\Omega|_{t=t_{peak}} \, \approx \, \frac{(C_1^v)^2+(C_1^w)^2}{C_1^wC_0^v - C_1^vC_0^w} \,\,\, .
\ee
We can also compute the width at half-height, defined by $\Delta t \equiv |t^*_1 - t^*_2|$ with $t^*_{1,2}$ being the two solutions of $\Omega (t^*) = \frac{1}{2} \,\Omega|_{t=t_{peak}}$\,. Using (\ref{Om_hid_sym_p}), we find:
\be
t^*_{1,2} \, \approx \, \frac{1}{\kappa} - \frac{C_1^vC_0^v+C_1^wC_0^w}{(C_1^v)^2+(C_1^w)^2} \pm \frac{(C_0^v C_1^w - C_1^v C_0^w)}{(C_1^v)^2+(C_1^w)^2} \,\,\, ,
\ee
which implies:
\be \label{delta_t_hid_sym}
\Delta t \, \approx \, \frac{2 |C_0^v C_1^w - C_1^v C_0^w|}{(C_1^v)^2+(C_1^w)^2} \,\,\, .
\ee
Notice, by comparing (\ref{Om_hid_sym_p}) and (\ref{delta_t_hid_sym}), that higher peaks of $|\Omega|$ are interrelated with smaller widths $\Delta t$ of the Gaussian-like shape of $\Omega (t)$\,. Finally, let us also note that, in the large-$u$ limit, (\ref{m2_uvw}) gives:
\be \label{MV_lu}
M^2_V = \frac{3V_0}{4} = \kappa^2 \,\,\, .
\ee

One can verify that the analytical estimates (\ref{t_peak_l}), (\ref{Om_hid_sym_p}) and (\ref{delta_t_hid_sym}) are very accurate by comparing them, for instance, to the numerical examples  plotted, by using the full expression (\ref{Om_uvw}), on Figures \ref{Tp_5} and \ref{Tp_10} in Appendix \ref{NumEx}. Furthermore, having analytical expressions for the characteristic features of $\Omega (t)$\,, clearly enables one to produce many more examples with $t_{peak} > 1$\,, as well as to vary the height and position of the peak as desired.\footnote{Note that there are six independent integration constants in (\ref{Sols_uvw})-(\ref{Constr_s}). So they can be expressed in terms of the six initial conditions $a_0 \equiv a|_{t=0}$\,, $\rho_0 \equiv \rho|_{t=0}$\,, $\theta_0 \equiv \theta|_{t=0}$\,, $\dot{a}_0 \equiv \dot{a}|_{t=0}$\,, $\dot{\rho}_0 \equiv \dot{\rho}|_{t=0}$\,, $\dot{\theta}_0 \equiv \dot{\theta}|_{t=0}$\,. Thus, we can view (\ref{Om_hid_sym_p}) as implying that the height of the turning rate peak, and thus the magnitude of the desired effect, depends on the choice of initial conditions. (Note, though, that the peak itself exists for every trajectory.) Of course, this simply reflects the fact that, by far, not all of our trajectories have sharp enough turns, although clearly infinitely many do. This is not a new feature of our model compared to \cite{PSZ,FRPRW}, or models relying on rapid turns in general. Nevertheless, this does not prevent the existence of rapid-turn attractors; see \cite{APR}. Of course, a more careful investigation of the phenomenologically-preferable initial conditions in our class of models could be informative and we hope to come back to it in the future.} In particular, in Appendix \ref{NumEx} we have shown various examples with values of $|\eta_{\perp}|_{t_{peak}}$ in a range suitable for PBH generation, according to \cite{FRPRW} (and references therein). We have also demonstrated that one can vary the number of e-folds $N(t_{peak})$ at will, as can be seen on Figures \ref{Tp_5} and \ref{Tp_10}. Note further, that (\ref{MV_lu}) is in perfect agreement with the small (by comparison to the peak) positive and constant value of $m_s^2$\,, before and after the period of tachyonic instability, that one finds in the numerical examples illustrated on those two figures. 

We should point out that, in the examples with $t_{peak} > 1$\,, the peak of the turning rate occurs at $\rho (t) <\!\!< 1$\,, as should be clear from (\ref{rho_uvw}) and (\ref{Approx_reg_0}). For instance, one has $\rho(t)|_{t_{peak}} \sim 10^{-9}$ in the examples on Figure \ref{Tp_5} (as illustrated on Figure \ref{R_Th_l_5}), whereas $\rho(t)|_{t_{peak}} \sim 10^{-16}$ in the examples on Figure \ref{Tp_10}. In other words, the field-space region of interest for us is a neighborhood of the center of the Poincar\'e disk, instead of its boundary as was the case for the original $\alpha$-attractors of \cite{KLR,KLR2,KL3,CKLR}. In this region, the $\varepsilon$ parameter of the solutions (\ref{Sols_asc})-(\ref{Constr_s}) is exceedingly small. (Indeed, for the examples on Figure \ref{Tp_5} one has $\varepsilon|_{t_{peak}} \sim 10^{-19}$\,, whereas for those on Figure \ref{Tp_10}: $\varepsilon|_{t_{peak}} \sim 10^{-32}$\,.) In fact, one can compute that it tends to zero as \,$\varepsilon \sim t^2 e^{-2\kappa t}$\,. So the slow roll approximation $\varepsilon <\!\!< 1$ is very well satisfied (and ever-improving with time).\footnote{Note that (\ref{Pot}) implies that the potential slow roll parameter $\varepsilon_V \!= \frac{1}{2} \left( \frac{\pd_{\varphi} V}{V} \right)^2 \rightarrow 0$ as $\varphi \rightarrow 0$\,, which is consistent with the Hubble $\varepsilon$-parameter's behavior discussed in the main text. Note also that, for small $\varphi$\,, the potential (\ref{Pot}) can be approximated as $V \approx V_0 \left[ 1 + 4 \tanh^2 \! \left( \frac{\sqrt{6}}{8}\,\varphi\right) \right]$\,, where we have used that $\cosh (2 \beta) = \frac{1+\tanh^2 \beta}{1-\tanh^2 \beta}$ for any $\beta$\,. To make a comparison to the standard $\alpha$-attractors, let us take $n=1$ and $\alpha = \frac{16}{9}$ inside equation (5.1) of \cite{KLR}, obtaining the potential $V_{KLR} = \tanh^2 \!\left( \frac{\sqrt{6}}{8} \,\varphi \right)$ for any $\varphi$\,. Hence, we have that $\varepsilon_V^{KLR} = \frac{3}{4 \,\sinh^4 \left( \frac{\sqrt{6}}{4}\,\varphi \right) }$\,, which implies that $\varepsilon_V^{KLR} \rightarrow 0$ for $\varphi \rightarrow \infty$\,, while $\varepsilon_V^{KLR}$ diverges for $\varphi \rightarrow 0$\,. This illustrates how, for the standard $\alpha$-attractors, slow roll occurs near the boundary of the Poincar\'e disk (corresponding to $\varphi \rightarrow \infty$), while moving inward in field space (i.e., toward smaller $\varphi$) leads to a gradual exit from the inflationary phase.} However, it turns out that the behavior of the parameter $\eta_{\parallel}$ is problematic phenomenologically. Namely, the rapid turn is preceded by a period with $\eta_{\parallel} \approx const > 1$\,, while after the turn $\eta_{\parallel}$ tends (very fast) to $3/2$ with increasing $t$\,. (Curiously, the value of $\eta_{\parallel}$ before the turn decreases with increasing $t_{peak}$\,, although $3/2$ appears to be a lower bound, as one can verify numerically.) So, although the solutions under consideration could be useful as descriptions of a brief PBH-generating period of non-slow roll inflation, they cannot, by themselves, represent realistic full-fledged inflationary models. 

Note that this conclusion is based as much on the behavior before the rapid turn, as on that after it. Indeed, although the cosmological scales, relevant for PBH generation, are much smaller than the CMB ones, there are still tens of e-folds needed (at the tail end of the observable 50-60 e-folds) of inflation after the turning-rate peak, as can be seen from the benchmark cases in \cite{FRPRW} (see also \cite{BHFSSS}, as well as the interesting discussion in \cite{KMRV}, albeit in the context of different kinds of models). Hence, it is important to have phenomenologically viable inflationary regimes both before and after the peak of the turning rate. We will see in the next Section how to achieve that. Note that a key role will be played by the large-$u$ limit introduced in this Section.

\section{New solution with modified $\eta$-parameter} \label{NewModSol}
\setcounter{equation}{0}

In the previous Section we showed that the exact solutions (\ref{Sols_asc})-(\ref{Constr_s}), obtained for $f(\varphi)$ and $V(\varphi)$ given by (\ref{f_D}) and (\ref{Pot}) respectively, have many desirable properties. However, the behavior of their parameter $\eta_{\parallel}$ is problematic phenomenologically. So in this Section we will aim at finding modified solutions, which do not have  such a problem. For that purpose, note that the hidden symmetry, used in \cite{ABL} to derive the solutions (\ref{Sols_asc})-(\ref{Constr_s}), played no role in Section \ref{ExactSol} here. Its only remnant (the corresponding conserved quantity) is a constant, $C^w_1$\,, which is on par with any of the other integration constants. So now we will relax the symmetry condition and modify (\ref{uvw_coord_tr}) suitably as a starting ansatz. Note that a generalized-coordinate transformation, like (\ref{uvw_coord_tr}), does not have to be adapted to a symmetry in order to be useful for simplifying the equations of motion, even if the simplification is to a lesser degree. We should also note that, incidentally, the analytical considerations of this Section will lead to improved understanding of the numerical results and the estimates of Section \ref{ExactSol}.

\subsection{Ansatz and solution}

Since now we are not imposing any symmetry condition, there is no reason to restrict the value of the $\alpha$-parameter in (\ref{f_al}). So we take the function $f$\,, determining the target space metric (\ref{Gmetric}), to have the general form for a Poincar\'e disk target, namely:
\be \label{f_q}
f (\varphi) \,= \,\frac{1}{q^2} \,\sinh^2 (q \varphi) \,\,\, ,
\ee
where we have denoted $q^2 = \frac{2}{3 \alpha}$ for convenience. Our goal in the following will be to simplify the equations of motion with a suitable generalized-coordinate transformation and, then, to find analytical solutions in a certain regime.

We begin by making the ansatz:
\bea \label{uvw_mod}
\tilde{u} &=& a^{\frac{1}{2p}} \,\cosh (q \varphi) \,\,\, , \nn \\
\tilde{v} &=& a^{\frac{1}{2p}} \,\sinh (q \varphi) \,\cos \theta \,\,\, , \nn \\
\tilde{w} &=& a^{\frac{1}{2p}} \,\sinh (q \varphi) \,\sin \theta \,\,\, ,
\eea
where $p$ is a positive constant. We also have more freedom in choosing the potential now. It is convenient to take:
\be \label{V_q}
V (\varphi , \theta) \,= \,V_0 \,\cosh^{6p} (q \varphi) \,\,\, .
\ee
Note that both (\ref{uvw_mod}) and (\ref{V_q}) reduce to their hidden symmetry counterparts, (\ref{uvw_coord_tr}) and (\ref{Pot}) respectively, for $p=\frac{1}{3}$ and $q=\sqrt{\frac{3}{8}}$\,. This was part of our motivation for those ansatze, in view of the relation between (\ref{f_q}) and (\ref{f_D}). In addition, with (\ref{uvw_mod}) and (\ref{V_q}), the behavior of the $\eta_{\parallel}$-parameter, as well as the related $\eta_V = \pd^2_{\varphi} V / V$ parameter, will improve significantly as we will see shortly.  

Let us now turn to the Lagrangian and equations of motion. For that purpose, note that the inverse of (\ref{uvw_mod}) is the transformation:
\bea \label{apt_mod}
a (t) &=& (\tilde{u}^2 - \tilde{v}^2 - \tilde{w}^2)^p \,\,\, , \nn \\
\varphi (t) &=& \frac{1}{q} \,{\rm arccoth} \!\left( \sqrt{\frac{\tilde{u}^2}{\tilde{v}^2 + \tilde{w}^2}} \,\right) \,\,\, , \nn \\
\theta (t) &=& {\rm arccot} \!\left( \frac{\tilde{v}}{\tilde{w}} \right) \,\,\, .
\eea
Substituting (\ref{f_q}), (\ref{V_q}) and (\ref{apt_mod}) in (\ref{L_class_mech}), we can see that the Lagrangian simplifies considerably when:
\be \label{qp}
q = \frac{1}{\sqrt{24}} \,\frac{1}{p} \,\,\, .
\ee
In that case, we find:
\be \label{Lagr_mod}
{\cal L} \,= \,12 p^2 \left( \tilde{u}^2 - \tilde{v}^2 - \tilde{w}^2 \right)^{3p-1} \!\left[ - \dot{\tilde{u}}^2 + \dot{\tilde{v}}^2 + \dot{\tilde{w}}^2 \right] \!- V_0 \,\tilde{u}^{6p} \,\,\, .
\ee

Motivated by the considerations of Section \ref{ExactSol}, we will look for solutions of the Euler-Lagrange equations of (\ref{Lagr_mod}) in the large-$u$ regime:\footnote{The present analogue of (\ref{Ineq_uvw_approx}) turns out to follow from (\ref{Approx_reg}), as will become clear below.}
\be \label{Approx_reg}
|\tilde{u}|,|\dot{\tilde{u}}| \,>\!\!> \,|\tilde{v}|, |\tilde{w}|, |\dot{\tilde{v}}|, |\dot{\tilde{w}}| \quad .
\ee
The $\tilde{u}(t)$ equation, in that regime, is:
\be \label{ut_EoM}
24 \,p \,\tilde{u} \,\ddot{\tilde{u}} + 24 \,p \,(3p-1) \,\dot{\tilde{u}}^2 - 6 \,V_0 \,\tilde{u}^2 \,= \,0
\ee
For convenience, let us take:
\be \label{u_mod_sol}
\tilde{u}(t) \,= \,C_u \,e^{k_u t} \,\,\, ,
\ee
which satisfies (\ref{ut_EoM}) with:
\be \label{Cond_u_sol}
k_u = \sqrt{\frac{V_0}{12}} \,\frac{1}{p} \,\,\, .
\ee
The Euler-Lagrange equations for $\tilde{v}$ and $\tilde{w}$, in the regime (\ref{Approx_reg}), are the same, namely:
\be \label{yt_EoM}
\ddot{\tilde{y}} + 2 (3p-1) k_u \dot{\tilde{y}} - (3p-1) k_u^2 \tilde{y} = 0 \,\,\, ,
\ee
where $\tilde{y} = \tilde{v}, \tilde{w}$\,. The two solutions of (\ref{yt_EoM}) are of the following form: $const_{\pm} \times e^{k_{\pm} t}$ with $k_{\pm}=- k_u [(3p-1) \pm \sqrt{(3p-1)3p}\,]$\,. So let us choose:
\be \label{vw_mod_sol}
\tilde{v} (t) = C_v e^{k_v t} \qquad {\rm and} \qquad \tilde{w} (t) = C_w e^{k_w t} \,\,\, ,
\ee
where
\bea \label{kvkw}
k_v &=& - k_u \left[ (3p-1) + \sqrt{(3p-1)3p} \,\right] \,\,\, , \nn \\ 
k_w &=& - k_u \left[ (3p-1) - \sqrt{(3p-1)3p} \,\right] \,\,\, .
\eea
Note that for large $p$ one has: $k_w/k_u = - (3p-1) + \sqrt{(3p-1)3p} \,\approx \,\frac{1}{2}$\,; in fact, this approximation is very good numerically for any $p>1$\,. Finally, evaluating the Hamiltonian $E_{\cal L}$\,, resulting from (\ref{Lagr_mod}), on the above solutions in their regime of validity, i.e. in the approximation (\ref{Approx_reg}), gives:
\be
E_{\cal L} \approx -12 p^2 \tilde{u}^{6p-2} \dot{\tilde{u}}^2 + V_0 \tilde{u}^{6p} = (C_u e^{k_u t})^{6p} \left[ V_0 - 12 p^2 k_u^2 \right] \,\,\, . 
\ee
So, due to (\ref{Cond_u_sol}), the Hamiltonian constraint $E_{\cal L} = 0$ is satisfied without the need to impose any additional relation between the integration constants.

Before we begin analyzing the turning rate and entropic mass, obtained from these new solutions, let us briefly comment on how the behavior of the parameter $\eta_{\parallel}$ is modified compared to the hidden symmetry case. We will discuss the function $\eta_{\parallel} (t)$ in more detail at the end, once we have established that $\Omega (t)$ and $m_s^2 (t)$ behave again in the desired manner. For now, note that, as in Section \ref{ExactSol}, the trajectories of the new solutions tend with time to the point $\varphi = 0$ (the minimum of the potential (\ref{V_q})) and their `large-$t$' behavior is approached very fast, since they are described by exponentials. Now, on solutions of the equations of motion, $\eta_{\parallel}$ is equal to the Hubble slow roll parameter $\eta_H\equiv - \frac{\ddot{H}}{2 H \dot{H}}$\,.\footnote{Indeed, by using (\ref{EinstEqs}) and (\ref{phi_0_dot}), one can verify that (\ref{eta_PP}) gives $\eta_{\parallel} = \eta_H$\,.} So one can easily find from the scale factor $a(t)$ in (\ref{apt_mod}) that, at large $t$, the $\eta_{\parallel}$-parameter tends to:
\be
\eta_{\parallel} = - \frac{\ddot{H}}{2 H \dot{H}} \, \approx \, \frac{3p - \sqrt{3p\,(3p-1)}}{2p} \,\,\, .
\ee
Note that for $p=\frac{1}{3}$ this gives exactly the behavior arising from the solutions (\ref{Sols_asc})-(\ref{Sols_uvw}) (while, in that case, equations (\ref{ut_EoM}) and (\ref{yt_EoM}) reduce exactly to the corresponding hidden symmetry ones), namely:
\be
\eta_ {\parallel}\, \approx \, \frac{3}{2} \,\,\, .
\ee
On the other hand, for any $p > 1$ the expression $\frac{3p - \sqrt{3p\,(3p-1)}}{2p}$ is well approximated numerically by $\frac{1}{4p}$\,; this approximation improves (and becomes very accurate) with increasing $p$\,.\footnote{For instance, for $p=3$ the exact expression gives $0.086$\,, while  $\frac{1}{4p}$ gives $0.083$\,.} So, for `large $p$', we can write:
\be \label{eta_par_mod}
\eta_{\parallel} \, \approx \, \frac{1}{4p} \,\,\, .
\ee
Clearly, this can be made as small as desired, thus ensuring the slow roll approximation $\eta_{\parallel} <\!\!< 1$\,, by choosing appropriately the value of $p$\,. Note that (\ref{eta_par_mod}) is perfectly consistent with the potential slow roll parameter $\eta_V$\,. Indeed, at large $t$\,, and thus small $\varphi$\,, one obtains from (\ref{V_q}):
\be
\eta_V = \frac{\pd_{\varphi}^2 V}{V} \, \approx \, 6 p q^2 = \frac{1}{4p} \,\,\, ,
\ee
where we have used (\ref{qp}). The agreement between $\eta_{\parallel}$ and $\eta_V$ is an affirmation that the slow roll approximation is well-satisfied. In view of the discussion in this paragraph, we will always assume from now on that $p>1$ for the new solutions (\ref{apt_mod}).

\subsection{Turning rate}

Let us now turn to the investigation of the turning rate $\Omega (t)$ for the new solutions. Substituting (\ref{f_q}), (\ref{V_q}) and (\ref{apt_mod}) into (\ref{Om_PD}), we find:
\be \label{Om_uvw_tilde_0}
\Omega \, = \, 6 V_0 p q^2 \,\frac{\tilde{u}^{6p-1} \left( \tilde{v} \dot{\tilde{w}} - \dot{\tilde{v}} \tilde{w} \right) \left( \tilde{u}^2 - \tilde{v}^2 - \tilde{w}^2 \right)^{\frac{3}{2} - 3p} }{\left[ (\tilde{v} \dot{\tilde{u}} - \dot{\tilde{v}} \tilde{u})^2 + (\tilde{w} \dot{\tilde{u}} - \dot{\tilde{w}} \tilde{u})^2 - (\tilde{v} \dot{\tilde{w}} - \dot{\tilde{v}} \tilde{w})^2 \right]} \,\,\, .
\ee
Note that, for $p=\frac{1}{3}$ and $q^2=\frac{3}{8}$\,, this expression has precisely the same form as (\ref{Om_uvw}). Substituting (\ref{qp}) in (\ref{Om_uvw_tilde_0}), we obtain:
\be \label{Om_uvw_tilde}
\Omega \, = \, \frac{V_0}{4p} \,\frac{\tilde{u}^{6p-1} \left( \tilde{v} \dot{\tilde{w}} - \dot{\tilde{v}} \tilde{w} \right) \left( \tilde{u}^2 - \tilde{v}^2 - \tilde{w}^2 \right)^{\frac{3}{2} - 3p} }{\left[ (\tilde{v} \dot{\tilde{u}} - \dot{\tilde{v}} \tilde{u})^2 + (\tilde{w} \dot{\tilde{u}} - \dot{\tilde{w}} \tilde{u})^2 - (\tilde{v} \dot{\tilde{w}} - \dot{\tilde{v}} \tilde{w})^2 \right]} \,\,\, .
\ee 

In the regime of validity of our $\tilde{u}$, $\tilde{v}$ and $\tilde{w}$ solutions, i.e. the large-$u$ limit (\ref{Approx_reg}), the expression (\ref{Om_uvw_tilde}) gives:
\be \label{Om_mod_lead}
\Omega \, = \, \frac{V_0}{4p} \, \frac{\tilde{u}^2 \left( \tilde{v} \dot{\tilde{w}} - \dot{\tilde{v}} \tilde{w} \right)}{\left[ (\tilde{v} \dot{\tilde{u}} - \dot{\tilde{v}} \tilde{u})^2 + (\tilde{w} \dot{\tilde{u}} - \dot{\tilde{w}} \tilde{u})^2 \right]} \,\,\, .
\ee
Substituting (\ref{u_mod_sol}) and (\ref{vw_mod_sol}), we then find:
\be \label{Om_t_expl}
\Omega (t) \, = \,\frac{V_0}{4p} \,\frac{ C_v C_w \,(k_w - k_v) \,e^{(k_v+k_w)t}}{\left[ C_v^2 \,(k_u - k_v)^2 \,e^{2k_vt} + C_w^2 \,(k_u - k_w)^2 \,e^{2k_wt}\right]} \,\,\, .
\ee
Note that, just as in Section \ref{ExactSol} (and the related Appendix \ref{NumEx}), the sign of $\Omega$ here is fixed by the values of the constants, determining the solutions, and cannot change with time. So, again, every trajectory has either $\Omega (t) \!< 0$ for $\forall t$ (angular motion in clockwise direction) or $\Omega (t) \!> \!0$ for $\forall t$ (angular motion in anti-clockwise direction).

The function (\ref{Om_t_expl}) has a single extremum at:
\be \label{t_peak_mod_0}
t_{peak} \, = \, \frac{1}{(k_w - k_v)} \,\ln \!\left( \bigg|\frac{C_v}{C_w}\bigg| \frac{(k_u - k_v)}{(k_u - k_w)} \right) \,\,\, ,
\ee
which is a maximum for the trajectories with $\Omega > 0$ and a minimum for the trajectories with $\Omega < 0$\,. Substituting (\ref{kvkw}) in (\ref{t_peak_mod_0}), we obtain:
\be \label{t_peak_mod}
t_{peak} \, = \, \frac{1}{2 \,k_u \sqrt{(3p-1)3p}} \,\, \ln \!\left[\bigg|\frac{C_v}{C_w}\bigg| \frac{\left( 3p+\sqrt{(3p-1)3p}\,\right)}{\left( 3p-\sqrt{(3p-1)3p}\,\right)} \right] \,\,\, .
\ee
It is, perhaps, useful to note that, at large $p$, this formula is well approximated by:\footnote{In fact, this approximation is very accurate already for $p=3$, in which case: $\ln (12 p)|_{p=3} = 3.58$\,, while \,$\ln \!\left[ \frac{3p+\sqrt{(3p-1)3p}}{3p-\sqrt{(3p-1)3p}} \right]\!\bigg|_{p=3} = 3.53$\,.}
\be
t_{peak} \, \approx \, \frac{1}{2 \,k_u \sqrt{(3p-1)3p}} \,\, \ln \!\left[ 12 p \,\bigg|\frac{C_v}{C_w}\bigg| \,\right] \,\,\, .
\ee 

Now, let us compute the magnitude of the turning rate at the extremum.
Substituting (\ref{t_peak_mod_0}) in (\ref{Om_t_expl}), we find:
\be
\Omega |_{t=t_{peak}} \, = \, \frac{V_0}{8p} \,\frac{C_v}{C_w} \bigg|\frac{C_w}{C_v}\bigg| \,\frac{(k_w - k_v)}{(k_u - k_v)(k_u - k_w)} \,\,\, .
\ee
Then, using (\ref{kvkw}) and (\ref{Cond_u_sol}) gives:
\be \label{Om_mod_peak}
\Omega |_{t=t_{peak}} \, = \,\frac{{\rm sgn}(C_v C_w) \,V_0}{12 \,k_u \,p^2} \,\sqrt{(3p-1)3p} \, = \, {\rm sgn}(C_v C_w) \,k_u\,\sqrt{(3p-1)3p} \,\,\, .
\ee
Note that, at large $t$\,, the expression (\ref{Om_t_expl}) behaves as \,$\Omega (t) \approx const \times e^{- 2 \,k_u \,\sqrt{(3p-1)3p} \,\,t}$\,. In other words, the turning rate tends fast to zero after the peak. So, to have a more symmetric shape of the function $\Omega (t)$ around the point $t=t_{peak}$\,, one needs to take smaller values of the ratio $|C_w/C_v|$\,, as $(k_u - k_v)^2 >\!\!> (k_u - k_w)^2$ for any $p > 1$ according to (\ref{kvkw}). Furthermore, smaller $|C_w/C_v|$ leads to larger $t_{peak}$\,, as can be seen from (\ref{t_peak_mod}).

Finally, we can also obtain an explicit formula for the width $\Delta t$ at half-height, defined by \,$\Delta t \equiv t^+_* - t^-_*$ \,with \,$t^{\pm}_*$ \,being the two solutions of:
\be \label{Omt*}
\Omega (t_*) = \frac{1}{2} \,\Omega(t_{peak}) \,\,\, ,
\ee
where \,$t_*^- < t_{peak}$ \,and \,$t_*^+ > t_{peak}$ \!\,. Substituting (\ref{t_peak_mod_0}) in (\ref{Omt*}), one finds:
\be
t_*^{\pm} \,= \,\frac{1}{(k_w-k_v)} \,\ln \!\left( \,\bigg| \frac{C_v}{C_w} \bigg| \frac{(k_u-k_v)}{(k_u-k_w)} \,(2\pm\sqrt{3}) \!\right) \,\,\, .
\ee
Therefore, the width is:
\be
\Delta t \,= \,\frac{1}{(k_w - k_v)} \,\ln \!\left( \frac{2 + \sqrt{3}}{2 - \sqrt{3}} \right) = \frac{1}{2 k_u \sqrt{(3p-1)3p}} \,\ln \!\left( \frac{2 + \sqrt{3}}{2 - \sqrt{3}} \right) \,\,\, .
\ee
Note that again, as in Section \ref{ExactSol}, higher peaks have smaller width.

\subsection{Effective entropic mass}

Now we turn to considering the entropic mass of the fluctuations around the new trajectories. For convenience, let us introduce separate notation for the two terms in the $M_V^2$ expression (\ref{MV}):
\be \label{m_sq_T1T2}
m_{T1}^2 \equiv \frac{f \dot{\theta}^2 \pd_{\varphi}^2 V}{(\dot{\varphi}^2 + f \dot{\theta}^2)} \qquad {\rm and} \qquad m_{T2}^2 \equiv \frac{\frac{f'}{2 f} \,\dot{\varphi}^2 \pd_{\varphi} V}{(\dot{\varphi}^2 + f \dot{\theta}^2)} \quad .
\ee

Substituting (\ref{f_q}) and (\ref{V_q})-(\ref{qp}) in (\ref{m_sq_T1T2}), we find:
\be \label{m2_T1}
m_{T1}^2 = \frac{V_0}{4p} \,\frac{\tilde{u}^{6p-2} (\tilde{u}^2 - \tilde{v}^2 -\tilde{w}^2)^{1-3p} \,[(6p-1)\tilde{v}^2 + (6p-1)\tilde{w}^2 + \tilde{u}^2] \,(\tilde{v}\dot{\tilde{w}} - \dot{\tilde{v}}\tilde{w})^2}{(\tilde{v}^2+\tilde{w}^2) \left[ (\tilde{v} \dot{\tilde{u}} - \dot{\tilde{v}} \tilde{u})^2 + (\tilde{w} \dot{\tilde{u}} - \dot{\tilde{w}} \tilde{u})^2 - (\tilde{v} \dot{\tilde{w}} - \dot{\tilde{v}} \tilde{w})^2 \right]}
\ee
and
\be \label{m2_T2}
\hspace*{-1cm}m_{T2}^2 = \frac{V_0}{4p} \,\frac{\tilde{u}^{6p} \left( \tilde{u}^2 - \tilde{v}^2 - \tilde{w}^2 \right)^{- 3p} \left[ \,\dot{\tilde{u}} (\tilde{v}^2 + \tilde{w}^2) - \tilde{u} (\tilde{v} \dot{\tilde{v}} + \tilde{w} \dot{\tilde{w}}) \right]^{\!2}}{(\tilde{v}^2+\tilde{w}^2) \left[ (\tilde{v} \dot{\tilde{u}} - \dot{\tilde{v}} \tilde{u})^2 + (\tilde{w} \dot{\tilde{u}} - \dot{\tilde{w}} \tilde{u})^2 - (\tilde{v} \dot{\tilde{w}} - \dot{\tilde{v}} \tilde{w})^2 \right]} \quad .
\ee
One can verify that for $p=\frac{1}{3}$ the sum of (\ref{m2_T1}) and (\ref{m2_T2}) has exactly the same form as $M_V^2$ in (\ref{m2_uvw}). 

In the regime of validity of the new solutions, namely approximation (\ref{Approx_reg}), the above two expressions become:
\be \label{m2_T1_lu}
m_{T1}^2 = \frac{V_0}{4p} \,\frac{ \tilde{u}^2 \,(\tilde{v}\dot{\tilde{w}} - \dot{\tilde{v}}\tilde{w})^2}{(\tilde{v}^2+\tilde{w}^2) \left[ (\tilde{v} \dot{\tilde{u}} - \dot{\tilde{v}} \tilde{u})^2 + (\tilde{w} \dot{\tilde{u}} - \dot{\tilde{w}} \tilde{u})^2 \right]}
\ee
and 
\be \label{m2_T2_lu}
m_{T2}^2 = \frac{V_0}{4p} \,\frac{\left[ \,\dot{\tilde{u}} (\tilde{v}^2 + \tilde{w}^2) - \tilde{u} (\tilde{v} \dot{\tilde{v}} + \tilde{w} \dot{\tilde{w}}) \right]^{\!2}}{(\tilde{v}^2+\tilde{w}^2) \left[ (\tilde{v} \dot{\tilde{u}} - \dot{\tilde{v}} \tilde{u})^2 + (\tilde{w} \dot{\tilde{u}} - \dot{\tilde{w}} \tilde{u})^2 \right]} \,\,\,\, .
\ee
Note that, in obtaining (\ref{m2_T1_lu}), we have assumed that $p \tilde{v}^2$ and $p \tilde{w}^2$ are negligible compared to $\tilde{u}^2$\,. Although for large enough $p$ this may not seem justifiable, one should keep in mind that the constant $C_u$\,, inside $\tilde{u}$\,, can always be taken sufficiently large to ensure it. Notice that $C_u$ cancels out of both (\ref{m2_T1_lu}) and (\ref{m2_T2_lu}) and, thus, its precise value does not affect the tachyonic instability. Now, adding (\ref{m2_T1_lu}) and (\ref{m2_T2_lu}), one finds:
\be \label{mV_tilde_lu}
m_V^2 \equiv m_{T1}^2 + m_{T2}^2 = \frac{V_0}{4p} \,\,\,\, ,
\ee
even without substituting the solutions for $\tilde{u}$, $\tilde{v}$ and $\tilde{w}$\,. This is the same result as for the large-$u$ limit of $m_V^2$ in the hidden symmetry case; see (\ref{MV_lu}).

Although we have obtained (\ref{mV_tilde_lu}) without using the explicit solutions, it is curious to note that they lead to a certain interplay between the functions $m_{T1}^2 (t)$ and $m_{T2}^2 (t)$\,. Indeed, let us substitute (\ref{u_mod_sol}) and (\ref{vw_mod_sol}) into (\ref{m2_T1_lu}) and (\ref{m2_T2_lu}). This gives:
\be \label{msT1_uvw}
m_{T1}^2 (t) = \frac{V_0}{4p} \,\frac{C_v^2 C_w^2 (k_w - k_v)^2 e^{2(k_v+k_w)t}}{(C_v^2 e^{2k_v t} + C_w^2 e^{2 k_w t})[C_v^2 (k_u - k_v)^2 e^{2 k_v t} + C_w^2 (k_u - k_w)^2 e^{2 k_w t}]}
\ee
and 
\be \label{msT2_uvw}
m_{T2}^2 (t) = \frac{V_0}{4p} \,\frac{[(k_u - k_v) C_v^2 e^{2 k_v t} + (k_u - k_w) C_w^2 e^{2 k_w t}]^2}{(C_v^2 e^{2 k_v t} + C_w^2 e^{2 k_w t}) [C_v^2 (k_u - k_v)^2 e^{2 k_v t} + C_w^2 (k_u - k_w)^2 e^{2 k_w t}]} \,\,\,\, .
\ee
From (\ref{msT1_uvw}) and (\ref{msT2_uvw}) one can see that $m_{T1}^2 (t)$ has a maximum at $t=t_{peak}$ and tends to zero at large $t$\,, whereas $m_{T2}^2 (t)$ has a minimum at $t=t_{peak}$ and tends to $\frac{V_0}{4p}$ at large $t$. Lastly, as a check on our computations, one can verify that the sum of (\ref{msT1_uvw}) and (\ref{msT2_uvw}) gives again (\ref{mV_tilde_lu}), as it should.

At this point, there is one final ingredient left to discuss in the entropic mass formula (\ref{m_s_f}), namely the $\varepsilon$ term. For that purpose, note that the Ricci scalar of the target space metric (\ref{Gmetric}), with $f(\varphi)$ given by (\ref{f_q}), is ${\cal R} = - 2 q^2 = - \frac{1}{12 p^2}$\,, where we have also used (\ref{qp}). Of course, as before, the important point is that ${\cal R} = const$ (and, furthermore, the value of this constant is smaller for larger $p$) and so, in the slow roll approximation $\varepsilon <\!\!< 1$\,, this term can be neglected. We will explain shortly why the $\varepsilon$-parameter is naturally small in the large-$u$ regime, without the need for any restrictions on the constants determining the characteristic features of $m_s^2 (t)$\,. 

Let us now summarize the above results for the behavior of the entropic mass. Combining (\ref{Om_mod_peak}) and (\ref{mV_tilde_lu}), we find that at the peak:
\be \label{m2_mod_peak}
m_s^2|_{t=t_{peak}} = (m_V^2 - \Omega^2)|_{t=t_{peak}} = - 3 k_u^2 p \,(3p-2) \,\,\,\, ,
\ee
where we have used (\ref{Cond_u_sol}). On the other hand, from (\ref{Om_t_expl}) and (\ref{mV_tilde_lu}) one can see that:
\be \label{m2_mod_as}
m_s^2 \, = \, m_V^2 - \Omega^2 \,\,\, \rightarrow \,\,\, 3 k_u^2 p \qquad \, {\rm for\,\,both} \, \qquad t \rightarrow 0 \,\,\,\, {\rm and} \,\,\,\, t \rightarrow \infty \quad .
\ee
So the effective mass-squared $m_s^2 (t)$ is negative at the peak of the turning rate $\Omega (t)$\,, for any $p>1$\,, and tends to a positive constant before and after a brief tachyonic period around $t=t_{peak}$\,. Note also that the magnitude of this transient instability increases for larger values of $p$\,.

\subsection{Slow roll parameters}

Let us now discuss in more detail the slow roll parameters for the new solutions obtained in this Section. Recall that, by definition, the $\varepsilon$-parameter is given by $\varepsilon = - \frac{\dot{H}}{H^2}$\,. Computing the expression, that follows from the scale factor in (\ref{apt_mod}), together with the solutions (\ref{u_mod_sol}) and (\ref{vw_mod_sol}) in the regime (\ref{Approx_reg}), we obtain:
\be \label{vep_mod}
\varepsilon (t) \, = \, \frac{(k_u-k_v)^2 C_v^2 e^{2k_vt} + (k_u-k_w)^2 C_w^2 e^{2k_wt}}{k_u^2 \,p \,C_u^2 e^{2k_ut}} \,\,\, .
\ee
Using (\ref{kvkw}) in (\ref{vep_mod}) and recalling that $p>1$\,, one can see that $\dot{\varepsilon} (t) < 0$ for $\forall t$\,. In other words, the function $\varepsilon (t)$ is smooth and monotonically decreasing. And, furthermore, it tends to zero exponentially fast with increasing $t$\,. On the other hand, at $t=0$ we have:
\be
\varepsilon|_{t=0} \, = \, \frac{(k_u-k_v)^2 C_v^2 + (k_u-k_w)^2 C_w^2}{k_u^2 \,p \,C_u^2} \,\,\, .
\ee
So by taking $C_u$ large enough, we can always ensure that $\varepsilon (t)$ is small at $t=0$\,, as well. That, then, guarantees that the slow roll condition $\varepsilon<\!\!<1$ is satisfied for any $t\ge 0$\,. Note that requiring large $C_u$ is consistent with ensuring that the approximation (\ref{Approx_reg}) is satisfied right from the start, i.e. at $t=0$\,. So the $\varepsilon$ parameter is necessarily (rather) small within the range of validity of the solutions (\ref{u_mod_sol}) and (\ref{vw_mod_sol}).

Now let us consider the $\eta_{\parallel}$ parameter. On solutions of the equations of motion, it coincides with $\eta_H \equiv -\frac{\ddot{H}}{2 H \dot{H}}$\,, as recalled above. So, using (\ref{apt_mod}), (\ref{u_mod_sol}) and (\ref{vw_mod_sol}), we find that the leading contribution in the regime (\ref{Approx_reg}) is:
\be \label{eta_mod}
\eta_{\parallel} (t) \, = \, \frac{1}{2 p k_u} \,\frac{[(k_u - k_v)^3 C_v^2 e^{2k_vt} + (k_u - k_w)^3 C_w^2 e^{2k_wt}]}{[(k_u - k_v)^2 C_v^2 e^{2k_vt} + (k_u - k_w)^2 C_w^2 e^{2k_wt}]} \,\,\, .
\ee 
Computing the derivative of this function, one can verify, even without using (\ref{kvkw}), that $\dot{\eta}_{\parallel} (t) < 0$ for $\forall t$ and everywhere in parameter space. Hence $\eta_{\parallel} (t)$ is monotonically decreasing with time. In fact, in view of (\ref{kvkw}), it is easy to realize that (\ref{eta_mod}) approaches fast its large-$t$ limit:
\be
\eta_{\parallel} \, \approx \, \frac{(k_u - k_w)}{2pk_u} = \frac{3p-\sqrt{(3p-1)3p}}{2p} \,\,\, .
\ee
As noted earlier, this is well-approximated by $\eta_{\parallel} \approx \frac{1}{4p}$ for any $p \gtrsim 2$\,. On the other hand, at $t \approx 0$\,, the expression (\ref{eta_mod}) acquires the form:
\be \label{eta_mod_t0}
\eta_{\parallel} \, \approx \, \frac{(k_u - k_v)}{2pk_u} = \frac{3p+\sqrt{(3p-1)3p}}{2p} \,\,\, ,
\ee
where we have used that $(k_u - k_v)^2 >\!\!> (k_u - k_w)^2$ for any $p>1$\,; we have also assumed that $|C_v| \gtrsim |C_w|$\,, in view of the discussion below (\ref{Om_mod_peak}).
Note that (\ref{eta_mod_t0}) is well-approximated by $\eta_{\parallel} \approx 3$ for any $p \gtrsim 2$\,. Hence, interestingly, we have found that, for large-enough $p$\,, the modified solutions of this Section describe a brief ultra-slow-roll inflationary period\footnote{Note, though, that the duration of this period can be varied at will, just as in Section \ref{ExactSol}.} followed by a long-term slow-roll expansion. This is reminiscent of the numerical results of \cite{BHFSSS}, where a temporary tachyonic instability occurs in certain two-field models due to the transition between two slow-roll phases of inflation. However, in \cite{BHFSSS} this transition does not result from a period of large turning rate, but instead from having separate potential terms for each of the two scalars. Thus, each scalar is driving a separate inflationary phase. In our case, on the other hand, only the field $\varphi$ enters the potential. 

Note that, as discussed above, both $\varepsilon (t)$ and $\eta_{\parallel} (t)$ are monotonically decreasing. In fact, after the turning-rate peak they tend fast, with increasing time, to the following: $\varepsilon \rightarrow 0$ and $\eta_{\parallel} \rightarrow const$. This is the same behavior as in the constant-roll solutions of \cite{MSY}.\footnote{Of course, the same remark applies to the solutions considered in our Section \ref{ExactSol}, although the slow roll approximation was violated there.} Just as in that reference then, to embed our solutions into a realistic model of the evolution of the Universe, one has to view them as valid up to a certain moment of time, after which a different effective description (possibly, including additional degrees of freedom) has to take over. This is a rather standard assumption in inflationary model-building as the exit from inflation (resulting in standard Big Bang evolution), although undoubtedly very important, is a separate and challenging topic of research by itself. We will comment more on the graceful exit issue at the end of the next Section.


\section{On PBH generation} \label{PBHgen}
\setcounter{equation}{0}

In Sections \ref{ExactSol} and \ref{NewModSol} we showed that the turning rate function of the solutions studied there has a Gaussian-like shape, perfectly in line with the starting assumption of \cite{FRPRW}. For the hidden symmetry solutions considered in Section \ref{ExactSol}, we also found that the height of the peak of the dimensionless turning rate $\eta_{\perp} (t) = \Omega / H$ can be varied at will, by choosing suitably the values of the integration constants. However, this is not the case for the modified solutions of Section \ref{NewModSol}, even though one can vary the magnitude of (\ref{m2_mod_peak}) compared to (\ref{m2_mod_as}) by changing $p$\,. The reason is that the Hubble parameter changes with $p$ as well. Indeed, to leading order in the regime (\ref{Approx_reg}), from (\ref{apt_mod}) and (\ref{u_mod_sol}) we have: $H \approx 2 k_u p$\,. Hence, using (\ref{Om_mod_peak}), we find that \,$\eta_{\perp}^2|_{t=t_{peak}} \approx \frac{3(3p-1)}{4p}$\,, which tends to $9/4$ for large $p$ and is always below that limit for any $p > 0$\,. This implies that, for PBH generation, one has to consider subleading corrections to the modified solutions, since we need $|\eta_{\perp}|_{t_{peak}} \approx 23$ according to the benchmark cases in \cite{FRPRW}. 
\begin{figure}[t]
\begin{center}
\hspace*{-0.2cm}
\includegraphics[scale=0.34]{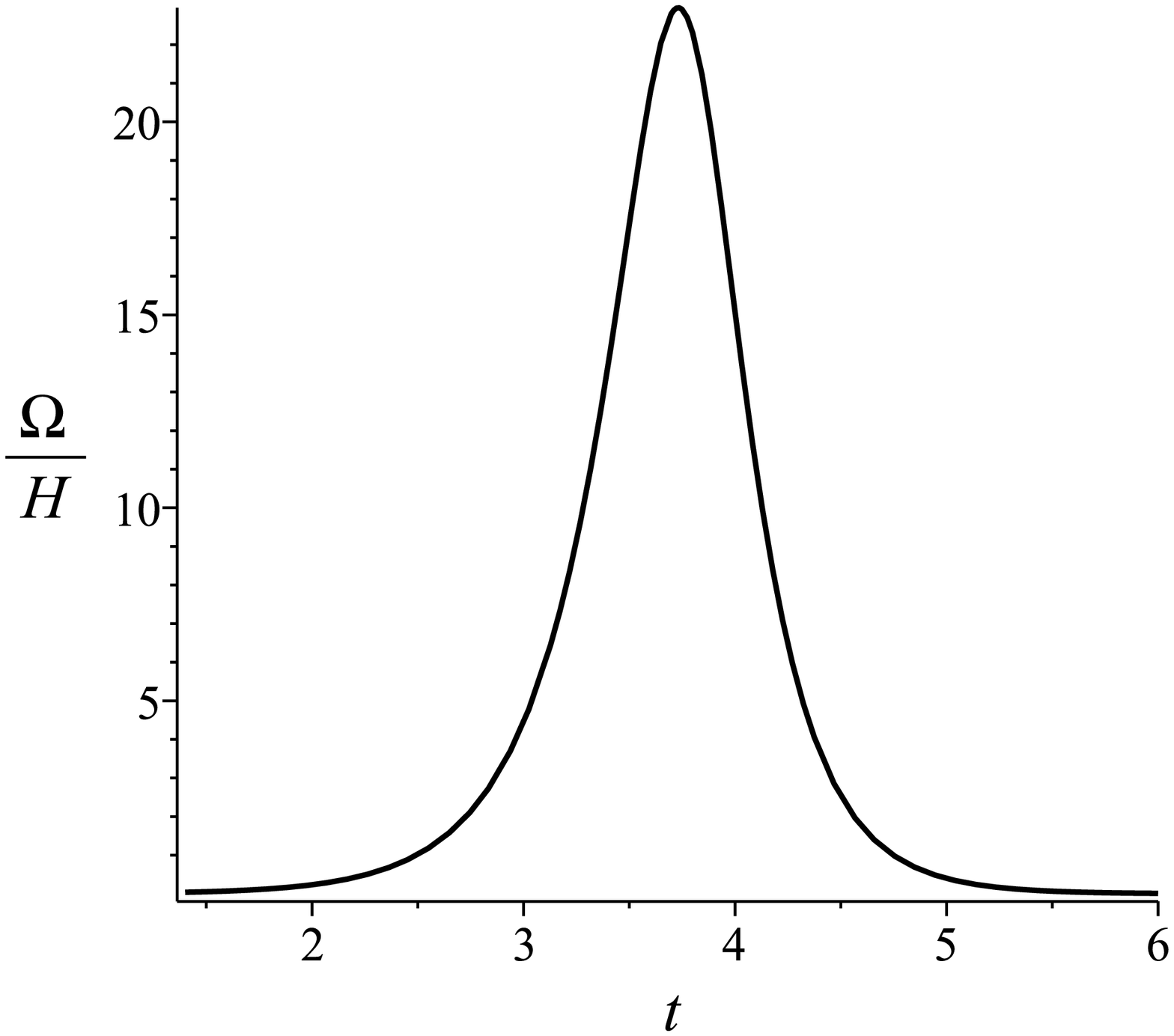}
\hspace*{0.5cm}
\includegraphics[scale=0.34]{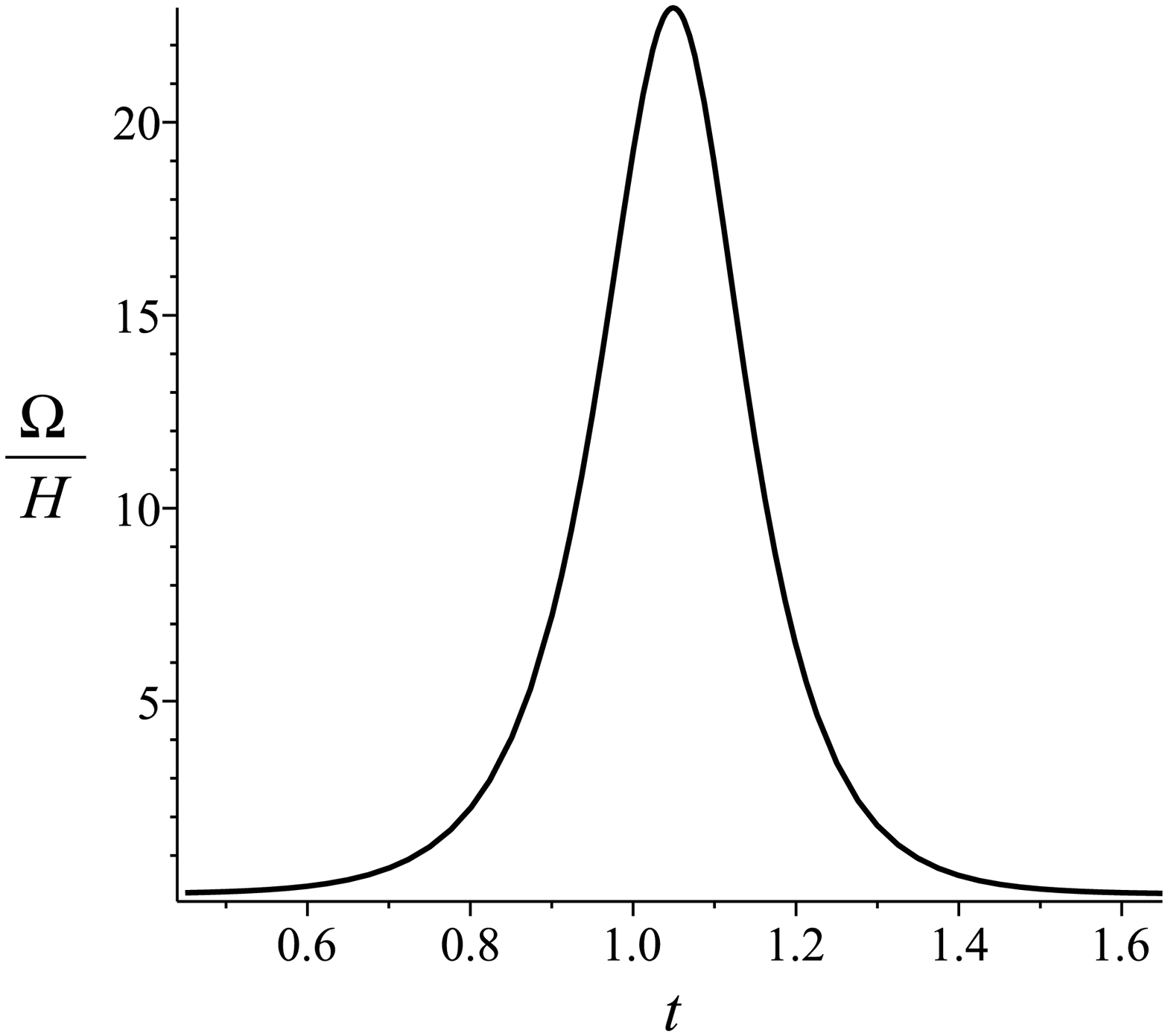}
\end{center}
\vspace{-0.7cm}
\caption{{\small The function $\eta_{\perp} (t) = \Omega / H$ for two examples with $N(t_{peak}) \approx 4$\,. On the left: $p=5$ and $C_w^{(1)} = 1/69$\,. On the right: $p=20$ and $C_w^{(1)} = 1/81$\,.  In both cases, the remaining constants are: $k_u = 1/10$\,, $C_v = 20$ and $C_w = 1/4$\,.}}
\label{OmH_mod_sol}
\vspace{0.1cm}
\end{figure}

Subleading corrections in $\tilde{v}(t)$ and $\tilde{w}(t)$\,, compared to (\ref{vw_mod_sol})-(\ref{kvkw}), can affect significantly the sharpness of the turn of a field-space trajectory (and thus the value of $|\eta_{\perp}|_{t_{peak}}$\,), while leaving the associated slow roll parameters essentially unchanged. This is because, as can be seen from (\ref{apt_mod}), the function $\dot{\theta} (t)$ is sensitive to such corrections, while $a(t)$ is not. Note that the modified solutions are approximate and that they improve numerically with time. Indeed, as is clear from (\ref{u_mod_sol}) and (\ref{vw_mod_sol})-(\ref{kvkw}), they satisfy the approximation (\ref{Approx_reg}), within which they were derived, to an ever-greater degree of accuracy with increasing $t$\,. So aiming to complete them to exact solutions (or, at least, to improve them), one may look for small corrections at (sufficiently) early times. We can write such corrections in the form:
\bea \label{Mod_cor}
\tilde{v}(t) &=& \left( \,C_v + C_v^{(1)} \,t + C_v^{(2)} \,t^2 + ... \,\right) e^{k_v t} \,\,\,\, , \nn \\
\tilde{w}(t) &=& \left( \,C_w + C_w^{(1)} \,t + C_w^{(2)} \,t^2 + ... \,\right) e^{k_w t} \,\,\,\, ,
\eea
where $C_{v,w}^{(1),(2),...} = const$\,. To exemplify the resulting effect, let us take for instance:
\be \label{Mod_cor_Cw_1}
\tilde{v}(t) \, = \, C_v \,e^{k_v t} \qquad {\rm and} \qquad \tilde{w}(t) \, = \, \left( C_w + C_w^{(1)} \,t \right) e^{k_w t}
\ee
with all other constants in (\ref{Mod_cor}) vanishing. Then, it is easy to obtain $\eta_{\perp} (t_{peak}) \approx 23$\,, as well as any other desired value, by choosing the constants carefully, as was also the case in Section \ref{ExactSol}; see (\ref{Om_hid_sym_p}). To illustrate this point, we have plotted on Figure \ref{OmH_mod_sol} the function $\eta_{\perp} (t) = \Omega / H$\,, obtained from substituting (\ref{Mod_cor_Cw_1}) in (\ref{Om_mod_lead}), for two examples with $N(t_{peak}) \approx 4$\,.\footnote{Note that, on the left of Figure \ref{OmH_mod_sol}, $H=1$ and thus $N(t)=\int H dt = t$, whereas on the right $H=4$ and so $N = 4t$\,. Hence, in both cases the sharp turn lasts about 3 to 4 e-folds, in line with \cite{FRPRW}.} We have taken the constants such that the $C_w^{(1)}$ contribution is a (reasonably) small correction for up to $t \approx t_{peak}$\,. One can notice that, for the example with greater $p$\,, the term \,$C_w^{(1)} t$ \,is a smaller correction. Of course, any putative corrections of the form (\ref{Mod_cor}) can only be viewed as small until a certain moment of time. We hope to report in the future on a more precise analytical investigation of subleading corrections to the modified solution of Section \ref{NewModSol}. 

The considerations of the present paper have shown that a fertile ground for producing field-space trajectories with a single (sharp) turn, which can realize the proposal of \cite{FRPRW} for a PBH-generating mechanism, is provided by (modifications of) a class of hidden symmetry solutions of \cite{ABL}. That class is characterized by having a Poincar\'e-disk scalar manifold and a rotationally-invariant scalar potential. It is worth investigating the properties of the turning rate for other classes (and appropriate modifications) of exact solutions of \cite{ABL}\footnote{As an interesting aside, it merits exploring whether (a suitable modification of) the considerations of \cite{ABL}, for the case of a hyperbolic-annulus scalar manifold and $\theta$-independent potential, could lead to a realization of the proposal of \cite{dAK} for rollercoaster cosmology, characterized by multiple alternating phases of accelerated and decelerated expansion.}, as well as for background solutions obtained by using more involved Noether symmetries, such as those in \cite{ABL2}. It would be very interesting to establish whether non-symmetric modifications are needed, in principle, to obtain a phenomenologically appropriate behavior of the Hubble $\eta$-parameter, in the cases with suitable $\Omega (t)$ functions. And if so, then whether symmetry-inspired ansatze, similar to the one in Section \ref{NewModSol} here, can lead to finding exact (albeit non-symmetric) background solutions with suitable for PBH-generation turning rates, in the more general context. 

It would also be very interesting to explore whether the present-case result, that slow roll occurs near the origin of field space\footnote{Note that (\ref{apt_mod}) implies for the radial coordinate on the disk: \,$\rho (t) = \frac{\sqrt{\tilde{v}^2 + \tilde{w}^2}}{\sqrt{\tilde{u}^2 - \tilde{v}^2 - \tilde{w}^2}+\sqrt{\tilde{u}^2}}$ \,, \,similarly to Section \ref{ExactSol}. So, in the regime (\ref{Approx_reg}), one always has $\rho <\!\!< 1$\,.} (unlike for standard $\alpha$-attractors), can be extended to broader classes of inflationary solutions in two-field models with hyperbolic scalar manifolds. Crucial for this conclusion was the form of the scalar potential that is required, or suggested, by the hidden symmetry. Hence, it is worth investigating the interrelation between the types of potentials, compatible with the more general Noether symmetries of \cite{ABL2}, and the kinds of field-space trajectories needed for PBH generation. In particular, that may enable us to find exact solutions, which combine the desirable properties of the solutions studied here (namely, small-field inflation and a brief rapid-turn period) with a graceful exit, that was lacking in our Sections \ref{ExactSol} and \ref{NewModSol}. In addition, the richer set of symmetry-compatible potentials in \cite{ABL2} might enable the existence of solutions with sharp turns that represent transitions between different pairs of inflationary regimes, compared to the ultra-slow-roll to slow-roll transition found here. Finally, in view of \cite{ASSV,APR}, it is also interesting to explore multi-field generalizations, with more than two scalars, of the kinds of solutions and field-space trajectories that we have investigated in the present work.

\section*{Acknowledgements}

I would like to thank E. M. Babalic, C. Lazaroiu, P. Suranyi and L.C.R. Wijewardhana for numerous discussions on various aspects of cosmological inflation, $\alpha$-attractors and primordial black holes. I am also grateful to J. Fumagalli for very useful correspondence. My work has received partial support from the Bulgarian NSF grants DN 08/3 and KP-06-N38/11.

\appendix

\section{Rescaling of hidden symmetry solutions} \label{Rescale}
\setcounter{equation}{0}

In Section \ref{ExactSol}, we have fixed from the start three of the constants in the exact background solutions (\ref{Sols_asc})-(\ref{Constr_s}), compared to the expressions in \cite{ABL}. In this Appendix we explain why this does not lead to any loss of generality.

The solutions for $a(t)$, $\varphi (t)$ and $\theta (t)$\,, that are of interest here, are given in (5.7) of \cite{ABL}, with $u(t)$, $v(t)$ and $w(t)$ as in (5.15), (5.18) and (5.21), together with the constraint (5.22), there. They contain three additional constants $C_1,C_2,C_w$ compared to (\ref{Sols_asc})-(\ref{Constr_s}) here. From \cite{ABL}, it is already clear that one can set $C_2 = 0$, without any loss of generality, since $C_2 \neq 0$ amounts only to mixing between $v$ and $w$. However, that still leaves, potentially, two additional constants. Namely, in the notation of Section \ref{ExactSol} here, we have the following expressions for the solutions \cite{ABL}:
\bea \label{Sols_asc_C1w}
a (t) &=& \left\{ u^2 - \left[ \frac{v^2}{C_1^2} + \frac{w^2}{C_w^2} \right] \right\}^{1/3} \,\,\, , \nn \\
\varphi (t) &=& \sqrt{\frac{8}{3}} \, {\rm arccoth} \!\left( \sqrt{\frac{u^2}{\frac{v^2}{C_1^2} + \frac{w^2}{C_w^2}}} \,\,\right) \,\,\, , \nn \\
\theta (t) &=& {\rm arccot} \!\left( \frac{C_w}{C_1} \,\frac{v}{w} \right) \,\,\, ,
\eea
where the functions $u (t)$ and $v (t)$ are the same as in (\ref{Sols_uvw}) while $w (t)$ has the form:
\be \label{Sols_w_Cw}
w (t) \,= \,C_w^2 \,C_1^w \,t + C_0^w
\ee
and, in addition, the constraint between the integration constants is:
\be \label{Constr_s_C1w}
(C_1^v)^2 + C_1^2 C_w^2 (C_1^w)^2 \,= \,\kappa^2 C_1^2 \!\left[ (C_1^u)^2 - (C_0^u)^2 \right] \,\,\, .
\ee
Note, though, that one can cancel $C_1$ and $C_w$ everywhere in (\ref{Sols_asc_C1w})-(\ref{Constr_s_C1w}) by performing the rescalings:
\be
C_1^w \rightarrow \frac{C_1^w}{C_w} \qquad , \qquad C_0^w \rightarrow C_w \,C_0^w \qquad , \qquad C_{0,1}^v \rightarrow C_1 \,C_{0,1}^v \quad ,
\ee
while keeping $C_{0,1}^u$ the same.

So, to recapitulate, without any loss of generality one can set:
\be \label{C1C2Cw}
C_1 = 1 \qquad , \qquad C_2 = 0 \qquad , \qquad C_w = 1
\ee
in the relevant solution of \cite{ABL}, thus obtaining (\ref{Sols_asc})-(\ref{Constr_s}) here.

\section{Radial variable of the exact solutions} \label{TheoremsRho}
\setcounter{equation}{0}

In this Appendix we prove that the radial coordinate $\rho (t)$ of the exact solutions (\ref{Sols_asc})-(\ref{Constr_s}) can have at most two local extrema. More precisely, we show that there are only three types of possible trajectories, namely: with no local extremum, with one local extremum (which is a maximum) and with two local extrema (a minimum followed in time by a maximum) of $\rho (t)$\,. 

To achieve this, let us analyze the condition $\dot{\rho} (t) = 0$ in detail. Substituting (\ref{Sols_uvw}) in (\ref{rho_extr_eq}), we find:
\bea \label{Eq_t_rho}
&&\hspace*{-0.7cm}\kappa \!\left\{  \!\left[ (C_1^v)^2 + (C_1^w)^2 \right] \!t^2 + 2 (C_1^v C_0^v + C_1^w C_0^w) t  +  \!\left[ (C_0^v)^2 + (C_0^w)^2 \right] \!\right\} \!\left[ C_0^u \sinh (\kappa \,t) \!+ \!C_1^u \cosh ( \kappa \,t) \right] \nn \\
&&\hspace*{-0.7cm}- \left\{ \left[ (C_1^v)^2 + (C_1^w)^2 \right] t + (C_1^v C_0^v + C_1^w C_0^w) \right\} \left[ C_0^u \cosh (\kappa \,t) \!+ \!C_1^u \sinh ( \kappa \,t) \right] = 0 \,\,\, .
\eea
Introducing the notation:
\bea \label{P1P2}
P_1 (t) &\equiv& \left[ (C_1^v)^2 + (C_1^w)^2 \right] t^2 + 2 \,(C_1^v C_0^v + C_1^w C_0^w) \,t  +  \left[ (C_0^v)^2 + (C_0^w)^2 \right] \,\,\, , \nn \\
P_2 (t) &\equiv& \left[ (C_1^v)^2 + (C_1^w)^2 \right] t + (C_1^v C_0^v + C_1^w C_0^w) \,\,\, ,
\eea
one can rewrite equation (\ref{Eq_t_rho}) as:
\be \label{rho_tanh}
\tanh (\kappa \,t) = Q (t) \,\,\, ,
\ee
where:
\be \label{RP12}
Q (t) \equiv \frac{C_0^u P_2(t) - \kappa \,C_1^u P_1(t)}{\kappa \,C_0^u P_1 (t) - C_1^u P_2 (t)} \,\,\, .
\ee
Since $\tanh (\kappa \,t)$ is a continuous and monotonic function, the key to determining the number of roots of (\ref{rho_tanh}), as well as estimating their positions, is the behavior of the function $Q(t)$\,. In the following we will study this behavior under the assumption that $C_{0,1}^u > 0$ [\,needed to ensure $a(t), \dot{a}(t) > 0$ for $\forall t$\,]\,, which implies in particular that $Q(t) \rightarrow - \frac{C_1^u}{C_0^u} < 0$ for $t \rightarrow \pm \infty$\,.

As for any rational function, the behavior of $Q(t)$ is determined by its extrema, singularities and zeros. Considering first the extrema, we find that $\dot{Q} = 0$ has only two roots:
\be \label{t_R_extr}
t_{\pm} = \frac{-C_0^v C_1^v - C_0^w C_1^w \pm |C_0^v C_1^w - C_1^v C_0^w|}{(C_1^v)^2+(C_1^w)^2} \,\,\, .
\ee
Note that the absolute value here ensures that $t_- \!< t_+$ always, regardless of the values of the constants. Now let us determine whether each of the two local extrema is a minimum or a maximum. For that purpose, we compute:
\be \label{Qdd}
\ddot{Q}|_{t=t_{\pm}} = \,\pm \,2 \,|C_0^v C_1^w - C_1^v C_0^w| \, \frac{\kappa \left[ (C_1^u)^2 - (C_0^u)^2 \right] [(C_1^v)^2+(C_1^w)^2]}{Q^2_{den} (t_{\pm})} \,\,\, ,
\ee
where $Q_{den}$ denotes the denominator of $Q(t)$ in (\ref{RP12}). Note that $\kappa > 0$ and $(C_1^u)^2 > (C_0^u)^2$ in our physical parameter space, as is evident from (\ref{Sols_uvw})-(\ref{Constr_s}). Hence, we always have $\ddot{Q} (t_-) < 0$ and $\ddot{Q} (t_+) > 0$\,; in other words, $t_-$ is always a local maximum, whereas $t_+$ is always a local minimum. Also, since $\dot{\theta} = \frac{C_0^v C_1^w - C_1^v C_0^w}{v^2+w^2}$\,, the requirement that $\dot{\theta}$ does not vanish identically means that we always have $C_0^v C_1^w \neq C_1^v C_0^w$ and thus $t_- \neq t_+$\,. 

Now we turn to the singularities of $Q(t)$, given by the zeros of its denominator. Solving $\kappa \,C_0^u P_1 (t) - C_1^u P_2 (t) = 0$\,, we obtain:
\be \label{t_s}
t^s_{\pm} = \frac{C_1^u}{2\kappa C_0^u} - \frac{(C_1^v C_0^v + C_1^w C_0^w)}{(C_1^v)^2 + (C_1^w)^2} \pm \left\{ \left( \frac{C_1^u}{2 \kappa C_0^u} \right)^{\!2} - \frac{(C_1^v C_0^w - C_1^w C_0^v)^2}{[(C_1^v)^2+(C_1^w)^2]^2} \right\}^{\!1/2} \,\, .
\ee
Clearly, $t^s_{\pm}$ is real only if (recall that $C_{0,1}^u > 0$):
\be \label{Real_ts}
\frac{C_1^u}{2 \kappa C_0^u} \ge \frac{|C_1^v C_0^w - C_1^w C_0^v|}{[(C_1^v)^2+(C_1^w)^2]} \,\,\, .
\ee
Notice that (\ref{t_R_extr}) implies in particular:
\be \label{t_pl_rel}
\frac{|C_1^v C_0^w - C_1^w C_0^v|}{[(C_1^v)^2+(C_1^w)^2]} = t_+ + \frac{(C_1^v C_0^v +C_1^w C_0^w)}{[(C_1^v)^2+(C_1^w)^2]} \,\,\, .
\ee
Combining this with (\ref{t_s}) and (\ref{Real_ts}), we conclude that:
\be
t_+ \le t^s_+ \,\,\, .
\ee
In other words, if $Q(t)$ has singularities, then both its extrema occur at earlier $t$ than the later-time singularity. Let us also compare $t_{\pm}$ to $t_-^s$\,. From (\ref{t_R_extr}) and (\ref{t_s}) we have:
\be
t_-^s \,= \,t_- + \frac{C_1^u}{2\kappa C_0^u} + \frac{|C_1^v C_0^w - C_1^w C_0^v|}{[(C_1^v)^2+(C_1^w)^2]} - \left\{ \left( \frac{C_1^u}{2 \kappa C_0^u} \right)^{\!2} - \frac{(C_1^v C_0^w - C_1^w C_0^v)^2}{[(C_1^v)^2+(C_1^w)^2]^2} \right\}^{\!1/2} \,\, .
\ee
Note that this is of the form $t_-^s = t_- + (p + q) - \sqrt{(p+q)(p-q)}$ with $p\ge q>0$\,. Since the inequality \,$(p + q) > \sqrt{(p+q)(p-q)}$ \,is true for any $p\ge q>0$\,, we conclude that:
\be
t_- < t_-^s \,\,\, .
\ee
Similarly, by using that \,$(p - q) < \sqrt{(p+q)(p-q)}$ \,is true for any $p\ge q>0$\,, we find:
\be \label{tms_tpl}
t_-^s < t_+ \,\,\, .
\ee

Let us now consider the zeros of $Q(t)$ in (\ref{RP12}). Computing the roots of the equation $C_0^u P_2(t) - \kappa \,C_1^u P_1(t) = 0$ gives:
\be \label{t_z}
t^z_{\pm} = \frac{C_0^u}{2\kappa C_1^u} - \frac{(C_1^v C_0^v + C_1^w C_0^w)}{(C_1^v)^2 + (C_1^w)^2} \pm \left\{ \left( \frac{C_0^u}{2 \kappa C_1^u} \right)^{\!2} - \frac{(C_1^v C_0^w - C_1^w C_0^v)^2}{[(C_1^v)^2+(C_1^w)^2]^2} \right\}^{\!1/2} \,\, .
\ee
So the condition for real $t^z_{\pm}$ is:
\be \label{Real_tz}
\frac{C_0^u}{2 \kappa C_1^u} \ge \frac{|C_1^v C_0^w - C_1^w C_0^v|}{[(C_1^v)^2+(C_1^w)^2]} \,\,\, .
\ee
This, together with (\ref{t_pl_rel}) and (\ref{t_z}), implies that:
\be \label{tpl_tplz}
t_+ \le t_+^z \,\,\, .
\ee
Notice also that, due to $C_1^u > C_0^u$\,, comparing (\ref{t_s}) and (\ref{t_z}) gives:
\be
t^z_+ < t^s_+ \,\,\, .
\ee
In addition, similarly to the comparison between $t_-$ and $t_-^s$\,, we find:
\be \label{tm_tmz}
t_- < t_-^z \,\,\, .
\ee
Note that the last inequality implies:
\be
t_-^s < t_-^z \,\,\, ,
\ee
since we have already shown that $t_-^s < t_+^z$\,; see (\ref{tms_tpl}) and (\ref{tpl_tplz}).\footnote{If it were true that $ t_-^z < t_-^s < t_+^z$\,, then one would need to have $t_-^z = t_-$ in contradiction with (\ref{tm_tmz}). That $t_-$ and $t_-^z$ cannot coincide in our parameter space follows also from the requirement $C_1^v C_0^w \neq C_0^v C_1^w$\,, needed to ensure that $\dot{\theta}$ is not identically zero; see the discussion below (\ref{Qdd}).}
\begin{figure}[t]
\begin{center}
\includegraphics[scale=0.6]{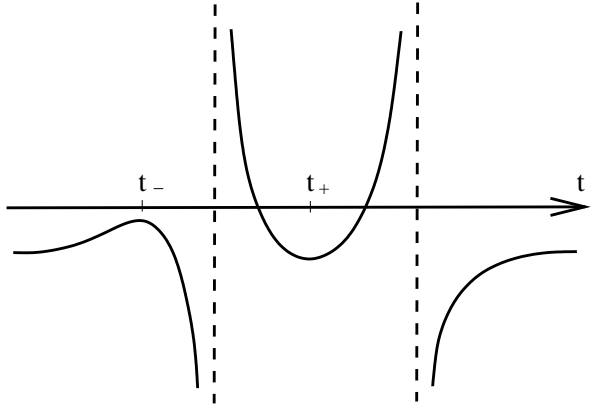}
\end{center}
\vspace{-0.7cm}
\caption{{\small Schematic depiction of the function $Q(t)$ in (\ref{RP12}), showing its characteristic features: two local extrema (a maximum at $t_-$ and a minimum at $t_+$), two singularities (at the positions of the vertical dashed lines) and two zeros, 
all situated with respect to each other according to the results derived in the text.}}
\label{RationalF}
\vspace{0.1cm}
\end{figure}

Let summarize our results so far regarding the behavior of the function $Q (t)$ in (\ref{RP12}). We have shown that it always has two local extrema situated at $t=t_{\pm}$\,, with $t_- \!< t_+$\,, and such that the extremum at $t_-$ is a maximum, while the extremum at $t_+$ is a minimum. Furthermore, when (\ref{t_s}) is real, $Q(t)$ also has two discontinuities, at the two singularities situated at $t=t_{\pm}^s$\,, such that $t_- < t_-^s < t_+ \le t_+^s$\,. And in addition, if (\ref{t_z}) is real, then $Q(t)$ has two zeros situated at $t = t_{\pm}^z$ and such that $t_-^s < t_-^z < t_+ \le t_+^z < t_+^s$\,. We have depicted schematically this general form of $Q(t)$ on Figure \ref{RationalF}, where we have also taken into account that $Q(t) \rightarrow - \frac{C_1^u}{C_0^u} < 0$ for $t \rightarrow \pm \infty$\,. Note that if there are zeros, then there must be singularities as well. However, it is possible to have singularities without having zeros.\footnote{Indeed, since $C_1^u > C_0^u$\,, the inequality (\ref{Real_tz}) implies (\ref{Real_ts}) but not vice versa.} In such a case, the middle segment (i.e., with $t_-^s<t<t_+^s$) is entirely above the horizontal axis. Finally, it is also possible to have neither zeros nor singularities, in which case the entire graph of $Q(t)$ is below the horizontal axis. 

It is remarkable that the above robust conclusions about the behavior of $Q(t)$ follow solely from the physical requirements that $\dot{\theta} (t)$ does not vanish identically and $a(t),\dot{a}(t)$ are positive-definite, regardless of any specific choices of parameter values. 

Now let us return to equation (\ref{rho_tanh}), whose solutions give the local extrema of the radial function $\rho (t)$ in (\ref{rho_uvw}), with (\ref{Sols_uvw}) substituted. Recall that we are interested in solving (\ref{rho_tanh}) in the interval $t \in [0,\infty)$\,, with $t=0$ being the initial moment of the expansion regime under study. In view of the general from of $Q(t)$, established here and depicted on Figure \ref{RationalF}, it is clear that the equation \,$\tanh (\kappa t) = Q (t)$ \,can have two, one or zero solutions in the interval $[0,\infty)$\,, depending on where the point $t=0$ is on the horizontal axis, compared to $t_{\pm}$\,, $t_{\pm}^s$ and $t_{\pm}^z$\,.\footnote{Note that if there are no singularities, then there are no solutions of (\ref{rho_tanh}) in the interval $[0,\infty)$\,, since the entire graph of $Q(t)$ is below the horizontal axis as explained above.} Hence, we have proven that the inflationary trajectories of the solutions (\ref{Sols_asc})-(\ref{Constr_s}) have at most two local extrema of $\rho (t)$\,, or equivalently of $\varphi (t)$\,. Note that the global minimum of the scalar potential (\ref{Pot}) is at $\varphi = 0$\,, i.e. at $\rho=0$\,. Therefore, all trajectories have to tend to $\rho = 0$ as $t \rightarrow \infty$\,. Hence, the last (or only) local extremum of $\rho(t)$ has to be a maximum. Denoting its position in time by $t_{max}$\,, we can conclude from the considerations of this Appendix that \,$t_+ \le t_{max} < t_+^s$\,. Furthermore, when $t_{\pm}^z$ is real, we have the following stronger restriction: \,$t_+^z < t_{max} < t_+^s$ \,.

\section{Numerical examples} \label{NumEx}
\setcounter{equation}{0}

In this appendix we give several numerical examples, which illustrate the possible types of  trajectories of the solutions (\ref{Sols_asc})-(\ref{Constr_s}), as well as their associated tachyonic instabilities. We will begin with some examples, that have $t_{peak} < 1$\,, for illustrative purposes. By the end, though, it will be abundantly clear that there are infinitely many examples with $t_{peak} > 1$ and thus with tens of e-folds occurring by the time $t=t_{peak}$ is reached. 

Before we start the numerical investigation, let us make several useful remarks regarding the physical constraints on the parameter space of our class of models. Of course, one constraint is the relation between constants, given in (\ref{Constr_s}), which in particular implies that $ |C_1^u| > |C_0^u|$\,. In addition, note from (\ref{uvw_coord_tr}) that to have real and positive-definite $a (t)$ and $\dot{a} (t)$ at any time, one needs $C_{0,1}^u > 0$\,. Another important restriction follows from the observation that, according to (\ref{Sols_asc})-(\ref{Sols_uvw}), the initial scale factor is $a(t)|_{t=0} = \left[ (C_0^u)^2 -(C_0^w)^2 - (C_0^v)^2 \right]^{\frac{1}{3}}$\,. Hence, to have $a(t)\ge 0$ for any $t\in[0,\infty)$\,, we need to impose: $(C_0^u)^2 -(C_0^w)^2 - (C_0^v)^2 \ge 0$ \,. Finally, note also that one cannot take simultaneously $C_0^v = 0$ and $C_0^w = 0$ for the following reason. If both of those constants were to vanish, then one would have $v/w = const$ and thus $\theta (t) \equiv const$. So we would be left with a single background scalar field $\varphi (t)$. However, here we are interested precisely in quantities and features of two-field models, like the turning rate, that distinguish them from single-field ones.   

\begin{figure}[t]
\begin{center}
\hspace*{-0.2cm}
\includegraphics[scale=0.33]{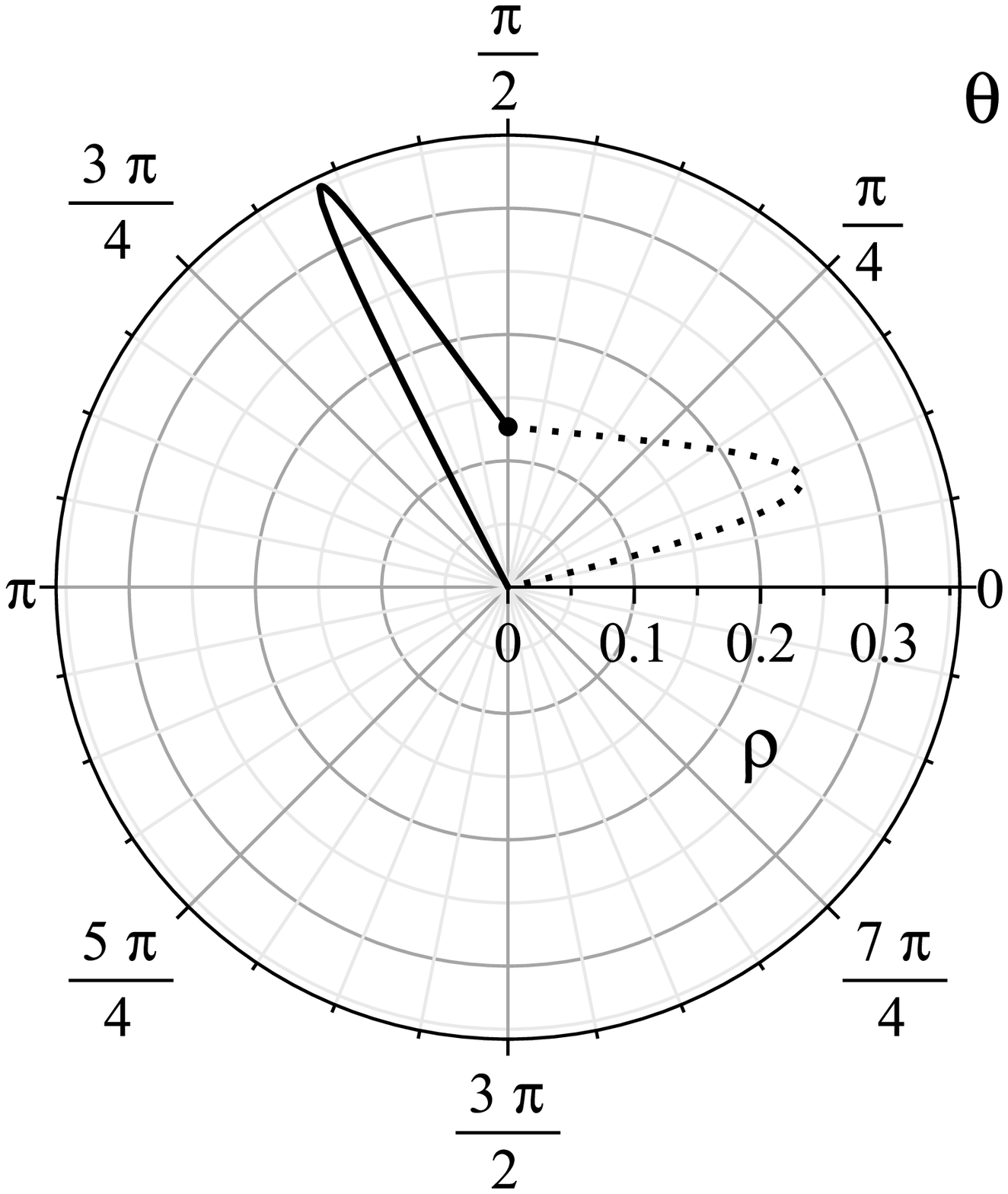}
\hspace*{0.5cm}
\includegraphics[scale=0.33]{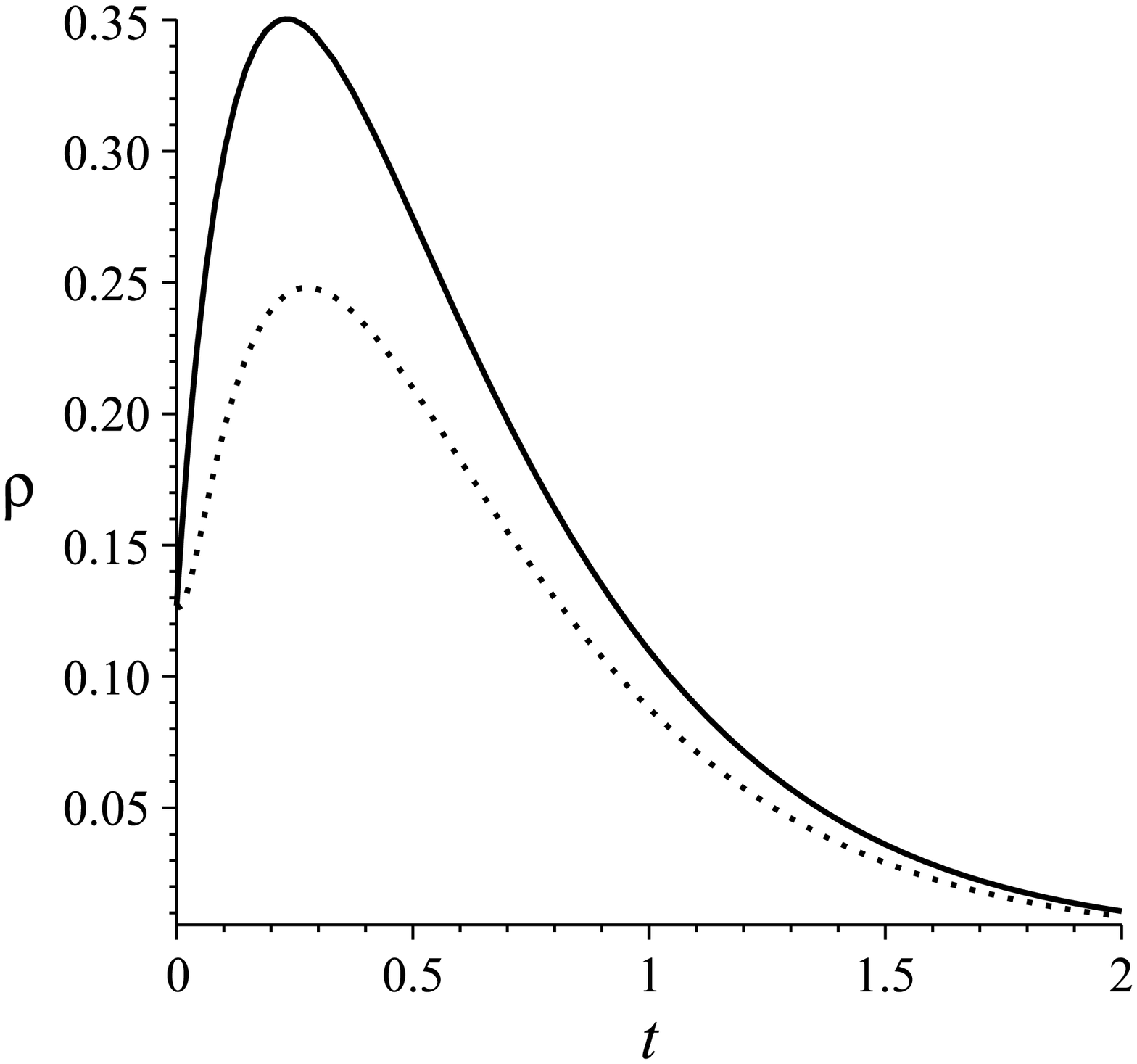}
\end{center}
\vspace{-0.7cm}
\caption{{\small Trajectories with a single local extremum of the radial function $\rho (t)$\,, one with positive-direction turning ({\it solid line}; $Ex1$ in the text), while the other with negative-direction turning ({\it dotted line}; $Ex2$ in the text). On the left: plot of the trajectories $\left( \rho (t) , \theta (t) \right)$ on the Poincar\'e disk. On the right: plot of their functions $\rho (t)$\,. The dot on the disk at $(\rho,\theta)=\left( \,0.13\,,\frac{\pi}{2} \,\right)$ 
denotes the starting point of the trajectories at $t=0$\,.}}
\label{RTh_Cv_0_Tps}
\vspace{0.4cm}
\end{figure}

\begin{figure}[t]
\begin{center}
\hspace*{-0.2cm}
\includegraphics[scale=0.33]{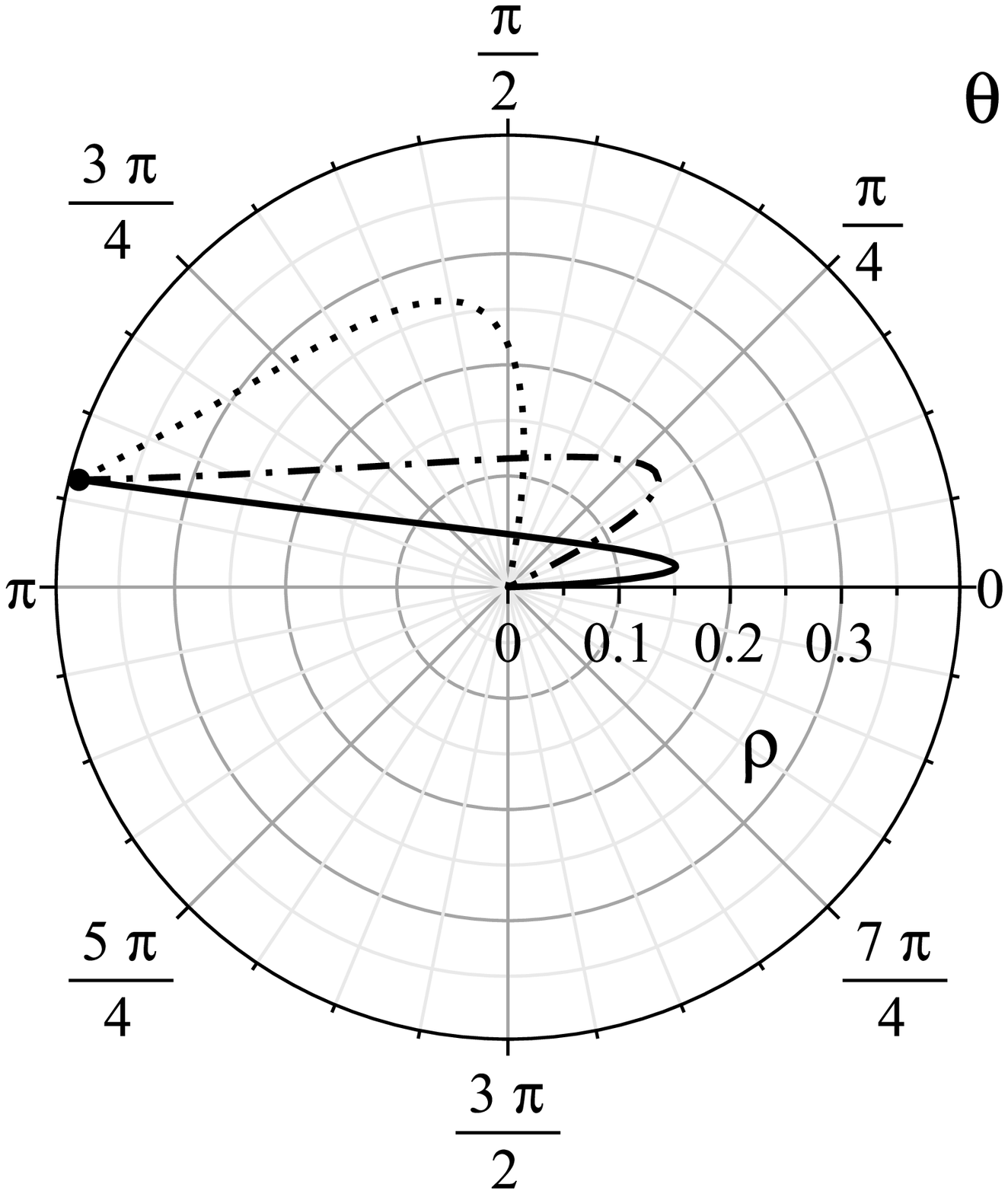}
\hspace*{0.5cm}
\includegraphics[scale=0.33]{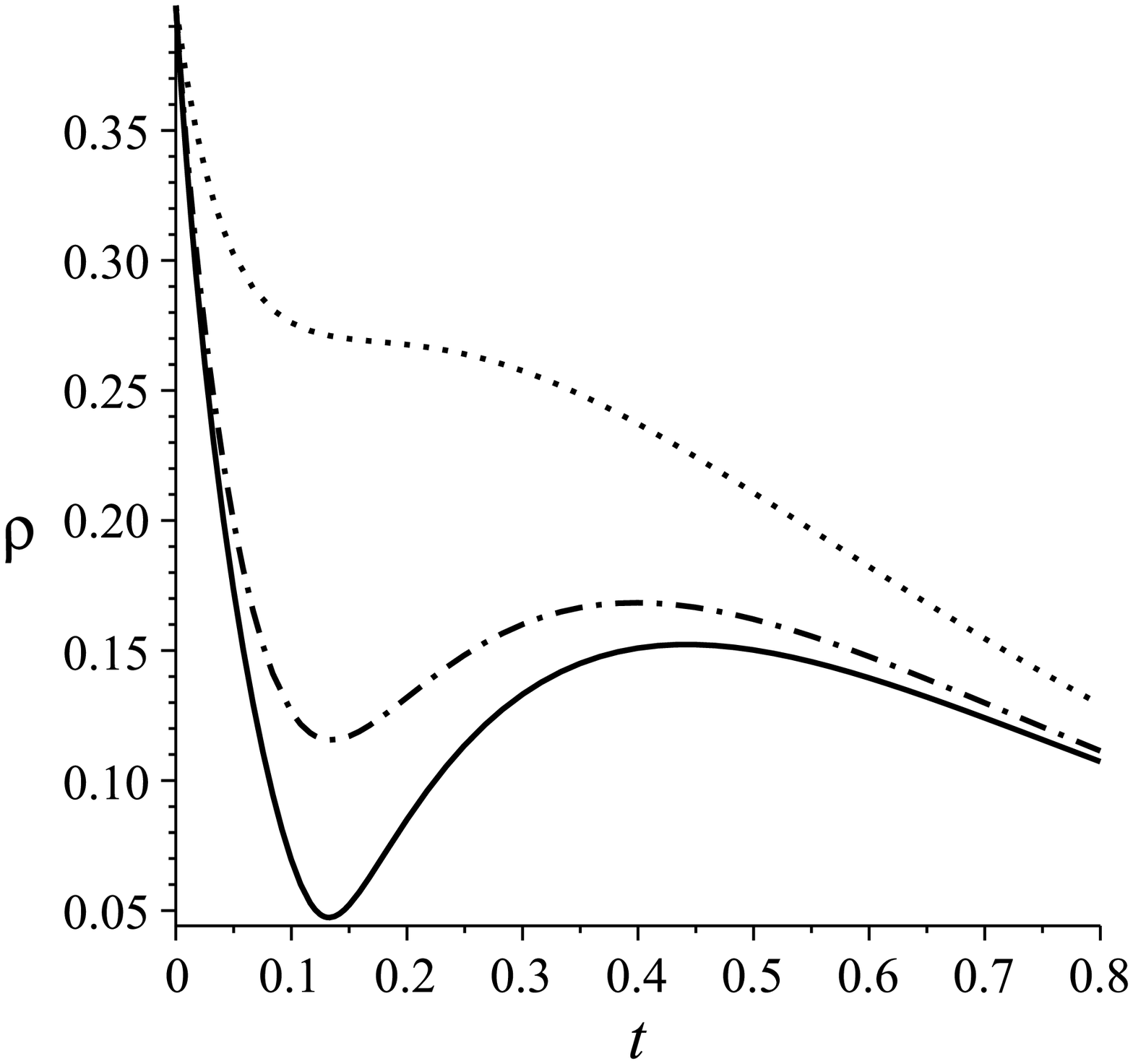}
\end{center}
\vspace{-0.7cm}
\caption{{\small Trajectories with two or with no local extrema of $\rho (t)$\,. On the left, the trajectories $\left( \rho (t) , \theta (t) \right)$ and, on the right, the functions $\rho (t)$ for examples $Ex5$ ({\it solid line}), $Ex6$ ({\it dash-dotted line}) and $Ex7$ ({\it dotted line}) in the text. The dot on the disk at $(\rho,\theta)\approx\left( \,0.4\,,\frac{15}{16} \,\pi \,\right)$ denotes the starting point of the trajectories at $t=0$\,.}}
\label{Cv_nRho}
\vspace{0.5cm}
\end{figure}

With the above comments in mind, let us now turn to numerical considerations, which will illustrate and supplement the analytical results of Appendix \ref{TheoremsRho}. In that appendix, we proved that 
the solutions (\ref{Sols_asc})-(\ref{Constr_s}) can have only three types of trajectories, namely with two, with one or with zero local extrema of the radial function $\rho (t)$\,. We show examples of all three types on Figures \ref{RTh_Cv_0_Tps} and \ref{Cv_nRho}; there, as well as in the rest of this Appendix, we have taken for convenience:
\be
\kappa = 3 \,\,\,.
\ee

\begin{figure}[h!]
\begin{center}
\hspace*{-0.2cm}
\includegraphics[scale=0.315]{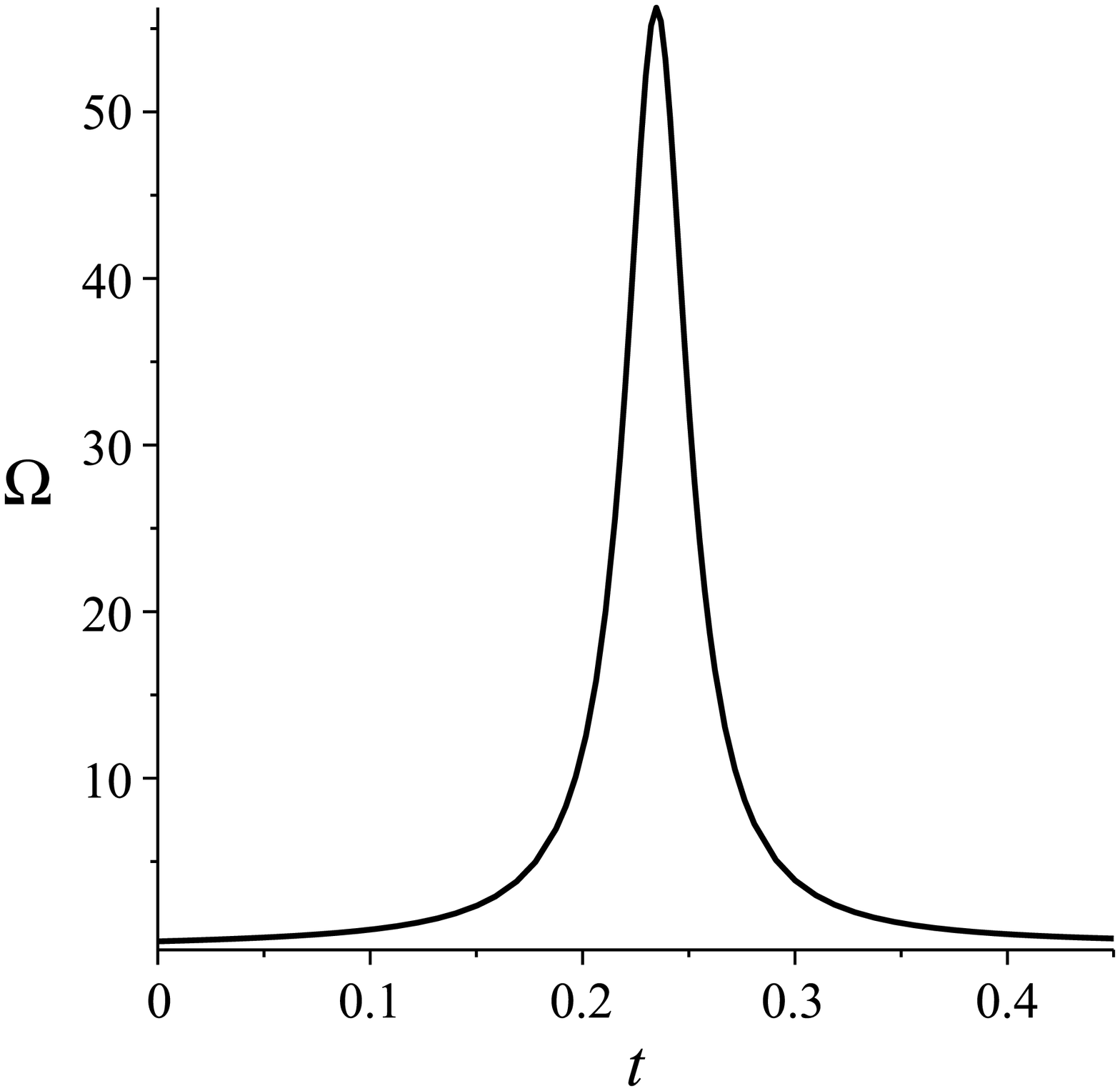}
\hspace*{0.5cm}
\includegraphics[scale=0.33]{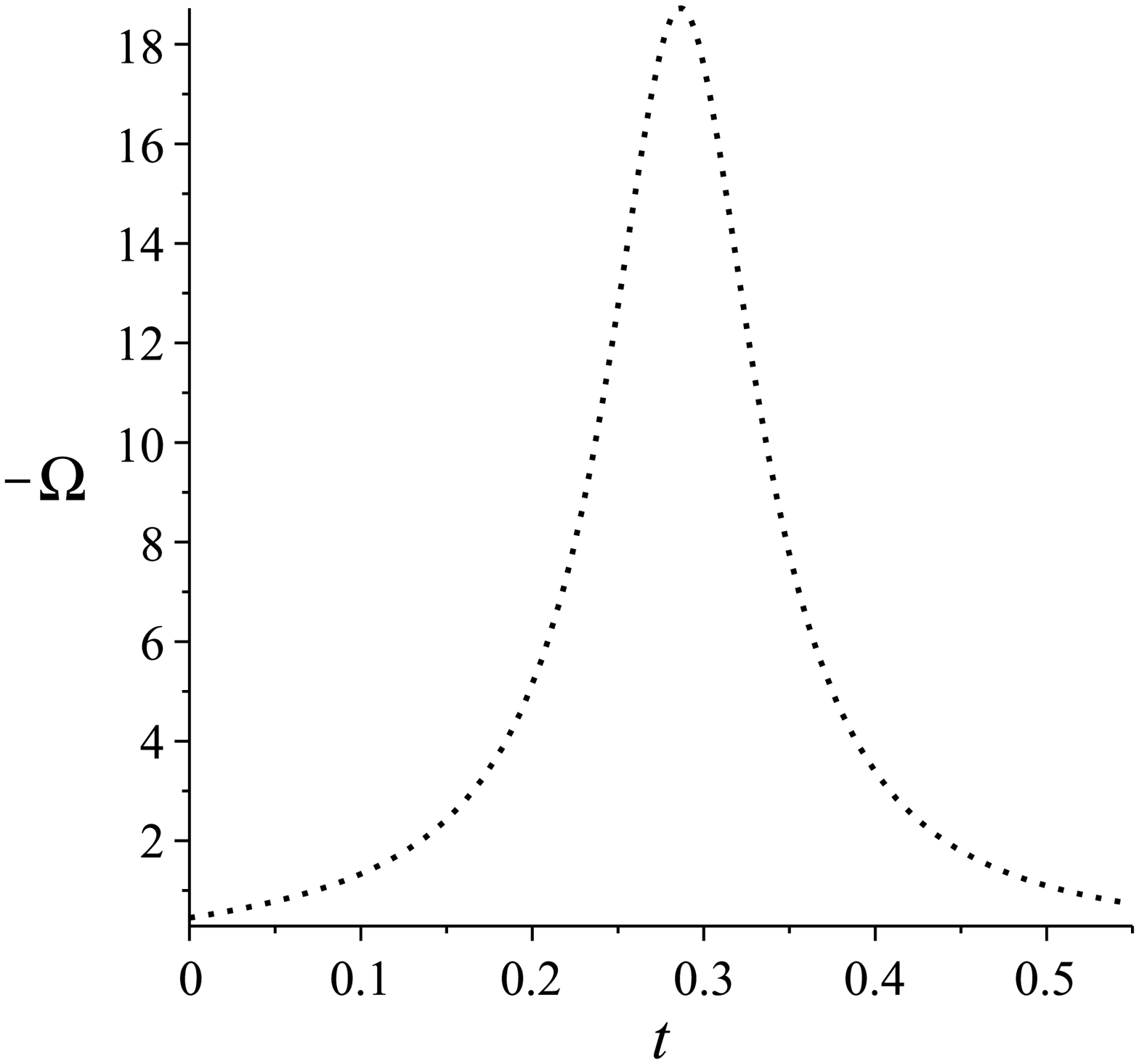}
\end{center}
\vspace{-0.7cm}
\caption{{\small The turning rate functions for the same trajectories as on Figure \ref{RTh_Cv_0_Tps}. As noted in the text, the sign of $\Omega (t)$ cannot change with time; it is fixed, for every trajectory, by the choice of integration constants.}}
\label{Om_Cv_0_Tps}
\vspace{0.1cm}
\end{figure}

\begin{figure}[t]
\begin{center}
\includegraphics[scale=0.34]{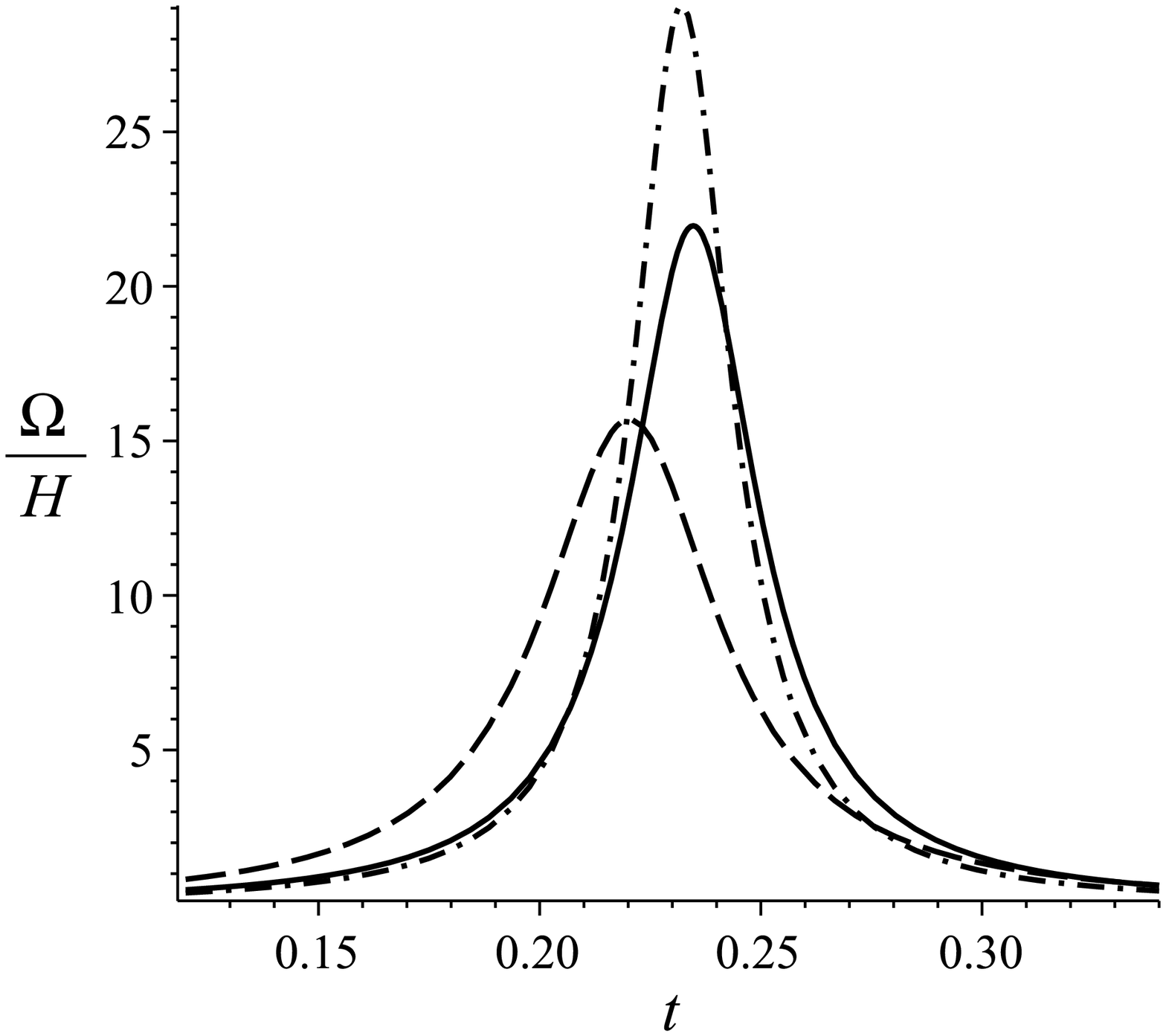}
\hspace*{0.4cm}
\includegraphics[scale=0.375]{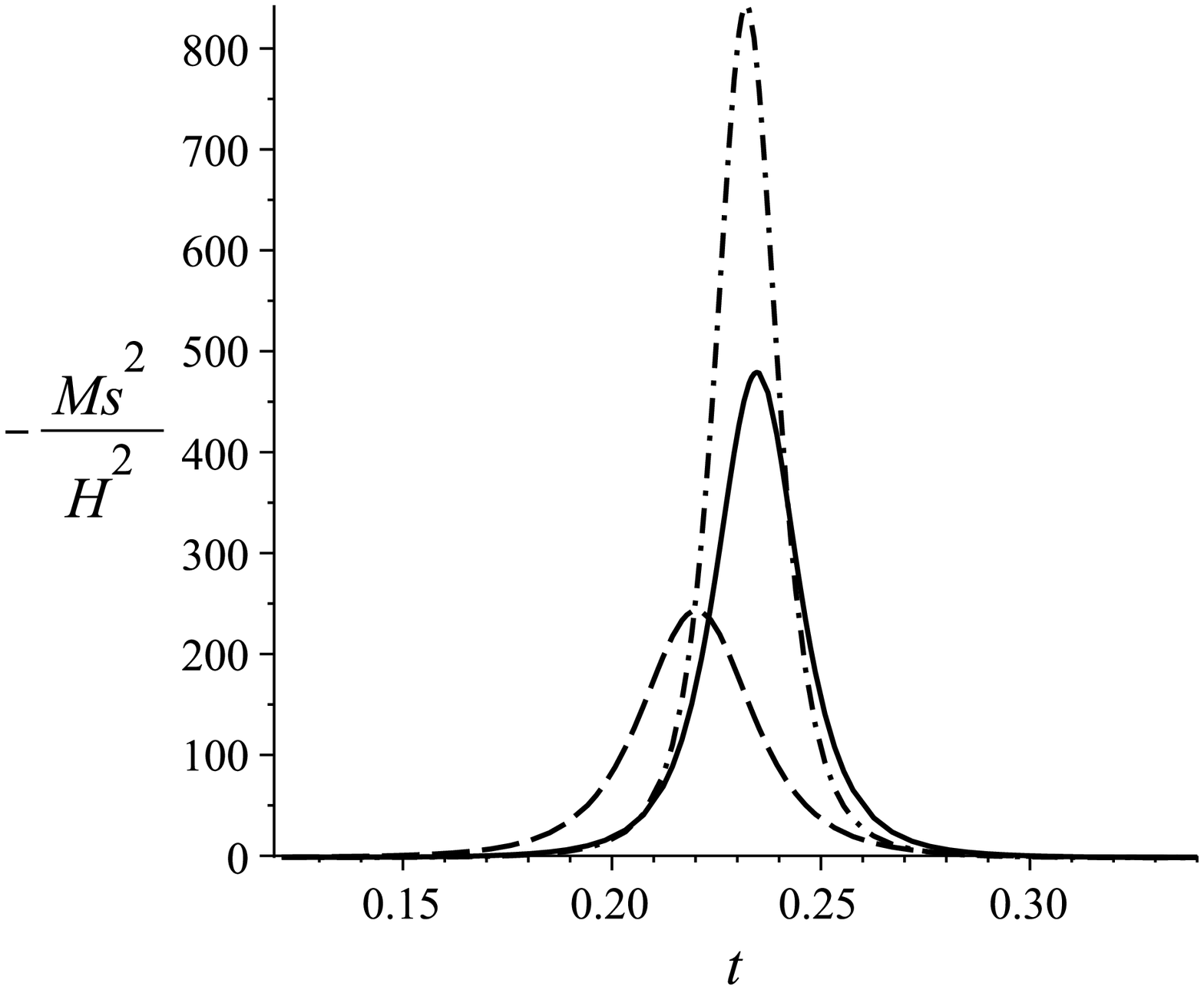}
\end{center}
\vspace{-0.7cm}
\caption{{\small The functions $\Omega/H$ (left) and $m_s^2/H^2$ (right) for three examples, demonstrating that the magnitude of the tachyonic instability can be varied as needed: the same $Ex1$ ({\it solid line}) as on Figure \ref{RTh_Cv_0_Tps}, as well as $Ex3$ ({\it dashed line}) and $Ex4$ ({\it dash-dotted line}) in the text. Note that, although not visible on the graphs, in each case $m_s^2/H^2$ has a small positive value outside of the brief period of instability.}}
\label{Om_H_Cv_0}
\vspace{0.1cm}
\end{figure}

On Figure \ref{RTh_Cv_0_Tps} we have plotted two trajectories, which have a single local extremum of $\rho(t)$, for the following sets of choices of the rest of the constants: $Ex1$ {\small ({\it solid line})}\,: $\{ C^u_1 = 3\,,  C^u_0 = 1\,, C^v_0 = 0\,, C^w_0 = \frac{1}{4}\,, C^v_1 = -4\,, C^w_1 = 2 \sqrt{14}\,\}$\,and $Ex2$ {\small ({\it dotted line})}\,: $\{C^u_1 = 2\,, C^u_0 = 1\,, C^v_0 = 0\,, C^w_0 = \frac{1}{4}\,, C^v_1 = \sqrt{26}\,, \,C^w_1 = 1\}$. Example $Ex1$ illustrates a trajectory with a positive-definite turning rate (note that the angular motion along the trajectory is in anti-clockwise direction), while example $Ex2$ illustrates a trajectory with $\Omega (t) < 0$ (i.e., angular motion in clockwise direction). We have shown the $\Omega(t)$ functions for these two trajectories on Figure \ref{Om_Cv_0_Tps}. Note that, since $C_{0,1}^u > 0$ as explained above, the sign of (\ref{Om_uvw}) is determined by the sign of $(v\dot{w}-\dot{v}w) = (C_0^v C_1^w - C_1^v C_0^w)$. Thus, the sign of $\Omega (t)$ is fixed by the choice of constants and cannot change with time.
On Figure \ref{Om_H_Cv_0} we show the dimensionless turning rate $\eta_{\perp} = \Omega/H$\,, as well as the tachyonic instability of the entropic mass $m_s^2/H^2$\,, (\ref{m_s_f}) with (\ref{Om_uvw})-(\ref{m2_uvw}) substituted, for the example $Ex1$ together with two other examples $Ex3$ and $Ex4$ obtained respectively by taking $C_0^w = \frac{1}{3}$ {\small ({\it dashed line})} and $C_1^v = -3$, $C_1^w = 3 \sqrt{7}$  {\small ({\it dash-dotted line})} with the rest of the constants as in $Ex1$\,. The point of examples $Ex3$ and $Ex4$ is to illustrate the ease with which one can vary the height of $\eta_{\perp}$\,, as well as the magnitude of the tachyonic instability, at $t \approx t_{peak}$\,. Note that, according to \cite{FRPRW}, a numerical value of \,$|\eta_{\perp}(t)|_{t=t_{peak}} \approx 23$ \,is needed to obtain the $10^7$ factor enhancement of the fluctuations power spectrum, necessary for PBH generation. This has guided our choices of constants in the examples given here. It should be noted, though, that one can obtain  any desired magnitude of the tachyonic instability by choosing suitably the values of the integration constants.

\begin{figure}[h!]
\begin{center}
\hspace*{-0.2cm}
\includegraphics[scale=0.34]{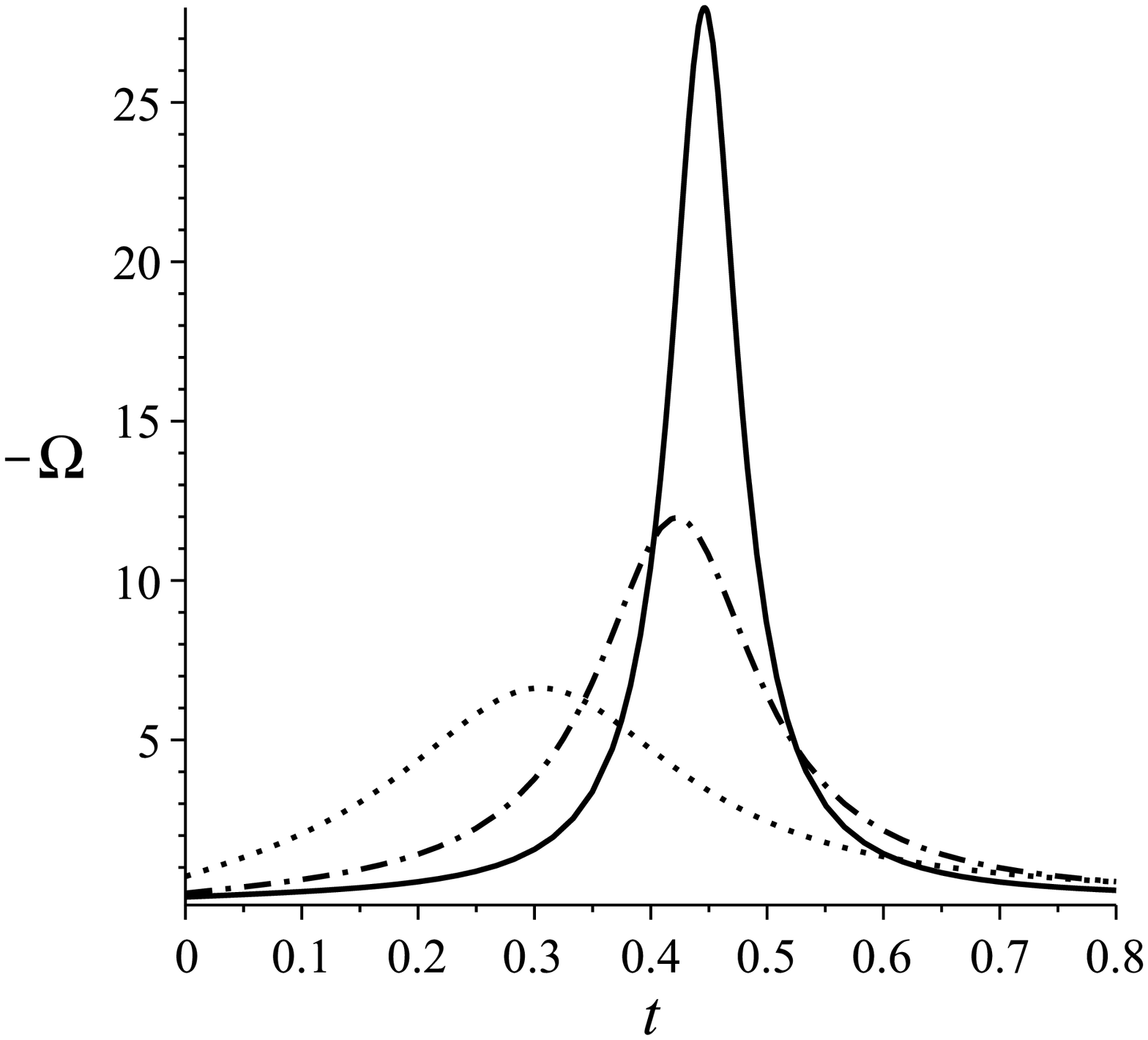}
\hspace*{0.4cm}
\includegraphics[scale=0.38]{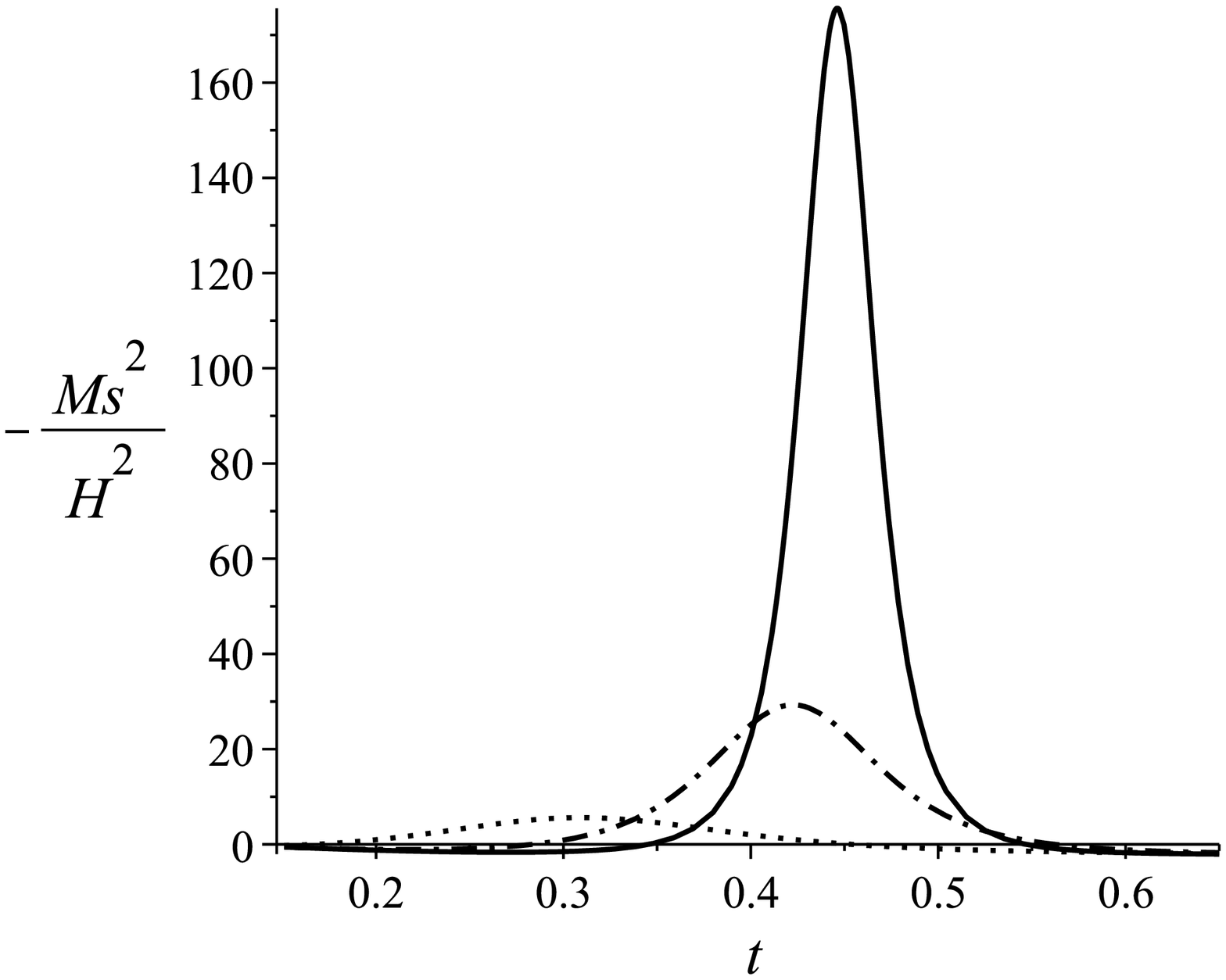}
\end{center}
\vspace{-0.7cm}
\caption{{\small The functions $-\Omega$ and $-m_s^2/H^2$ for the same trajectories as on Figure \ref{Cv_nRho}. Notice that $m_s^2/H^2$ has a small positive value outside of the brief tachyonic instability period.}}
\label{Cv_n_Om_Ms}
\vspace{0.1cm}
\end{figure}

\begin{figure}[t]
\begin{center}
\hspace*{-0.2cm}
\includegraphics[scale=0.35]{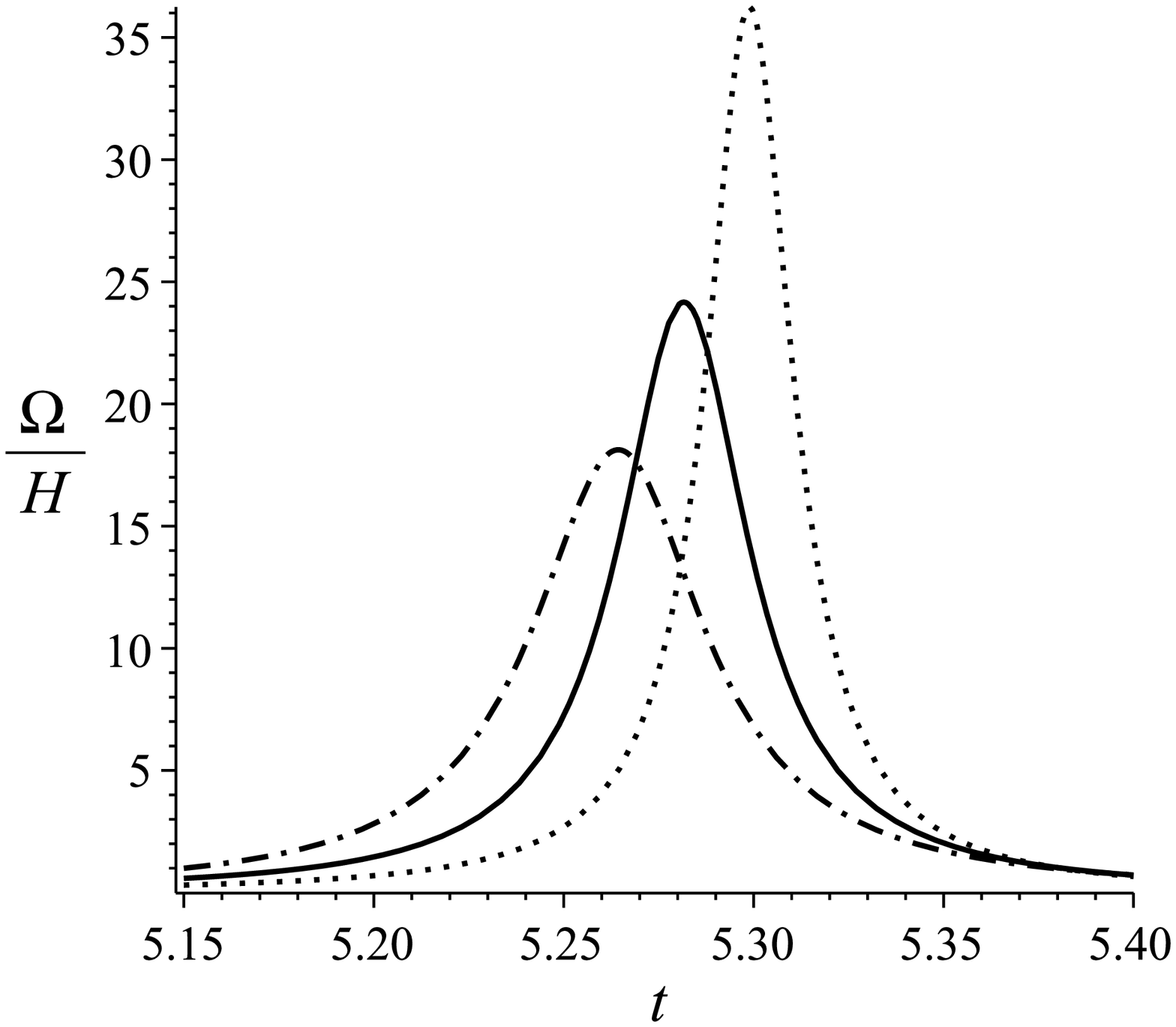}
\hspace*{0.4cm}
\includegraphics[scale=0.38]{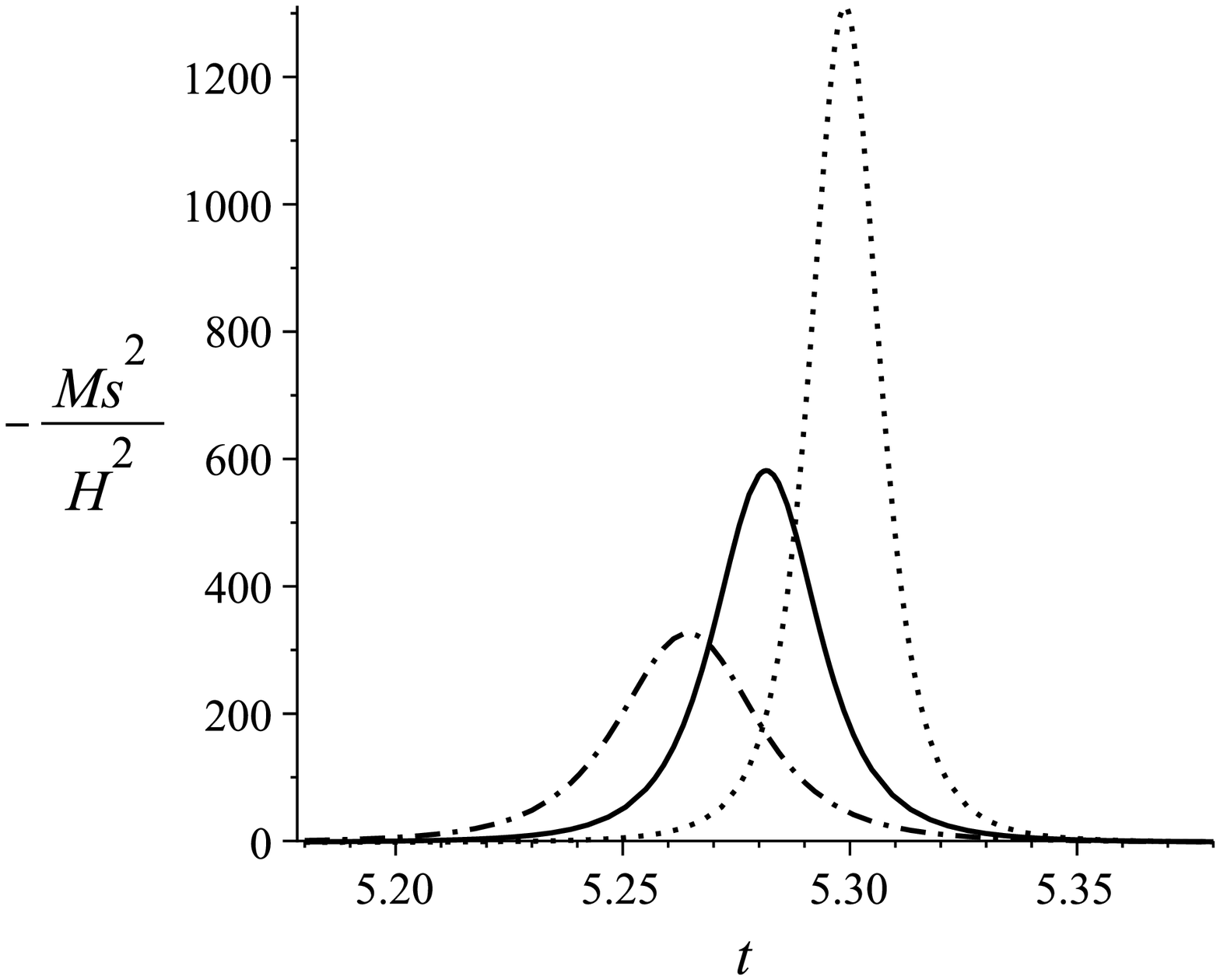}
\end{center}
\vspace{-0.7cm}
\caption{{\small The functions $\Omega/H$ and $-m_s^2/H^2$ for examples $Ex8$ ({\it dash-dotted line}), $Ex9$ ({\it solid line}) and $Ex10$ ({\it dotted line}) in the text. In these examples the number of e-folds by the time the peak is reached is $N(t_{peak}) = 11$\,. Note that, again, $m_s^2$ is small and positive outside of the tachyonic period, although this is not visible on the graphs.}}
\label{Tp_5}
\vspace{0.5cm}
\end{figure}
\begin{figure}[h!]
\begin{center}
\includegraphics[scale=0.315]{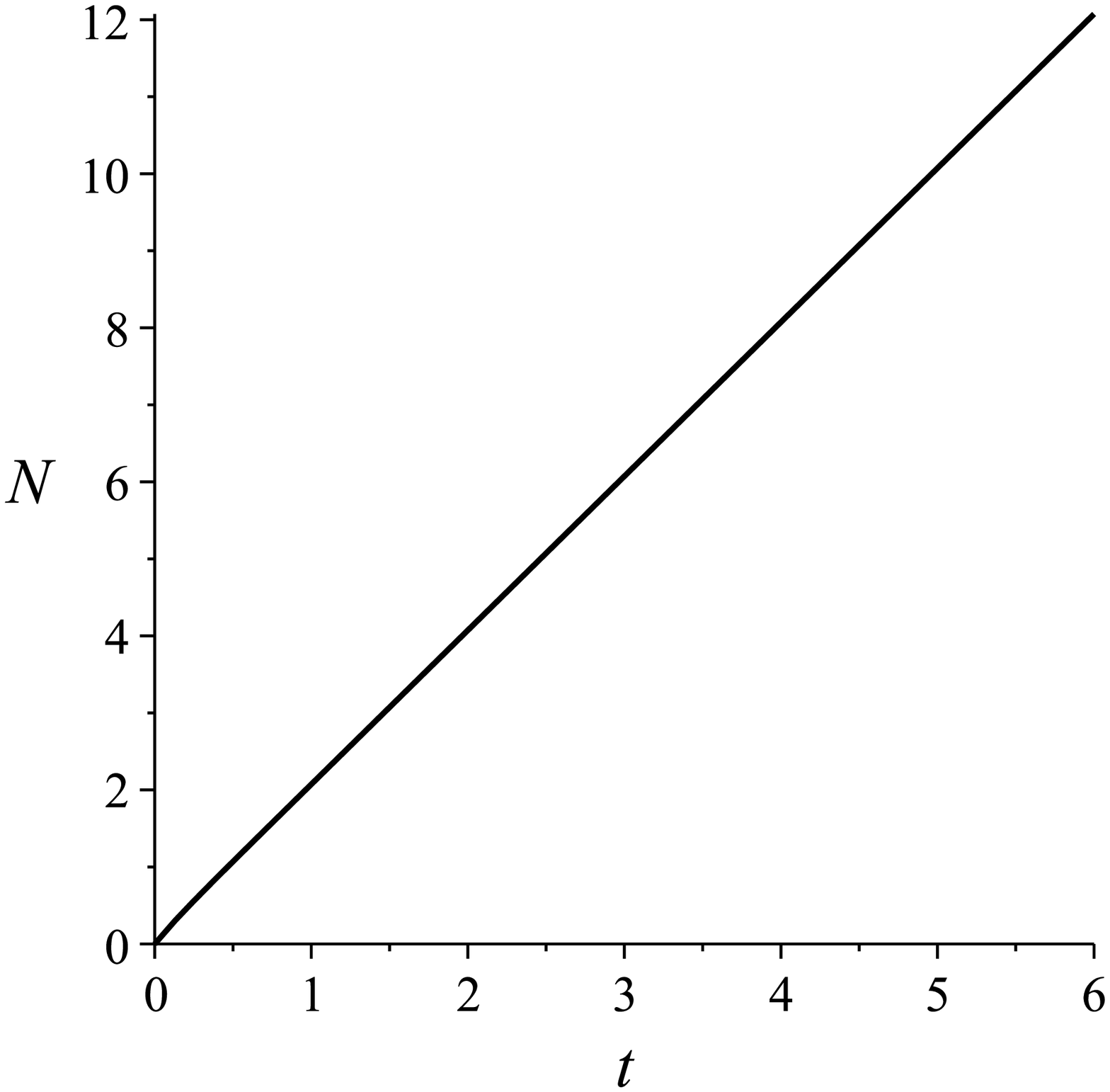}
\end{center}
\vspace{-0.7cm}
\caption{{\small The number of e-folds $N$, as a function of $t$\,, for the same examples as on Figure \ref{Tp_5}. The three graphs of $N(t)$, corresponding to the three examples, coincide on this plot.}}
\label{N_5}
\vspace{0.1cm}
\end{figure}

On Figure \ref{Cv_nRho} we illustrate the types of trajectories, which have either two or zero local extrema of $\rho (t)$\,. More precisely, we have plotted three trajectories, examples $Ex5$, $Ex6$ and $Ex7$ respectively, for the following choices of constants: $C^u_1=3$ , $C^u_0=\frac{3}{2}$ , $C^v_0=-1$ and $C^w_0=\frac{1}{4}$ together with $C^w_1=\frac{1}{8}$ {\small ({\it solid line})} , $C_1^w = 3$ {\small ({\it dash-dotted line})} and $C_1^w = 7.5$ {\small ({\it dotted line})}\,; in all three cases we have taken the value of $C_1^v$\,, determined by the positive root of (\ref{Constr_s}). On Figure \ref{Cv_n_Om_Ms} we have plotted the turning rate and resulting entropic mass for the same three examples. Together, Figures \ref{Cv_nRho} and \ref{Cv_n_Om_Ms} illustrate the relation between the shape of a trajectory and the height of its turning-rate peak, as well as the magnitude of the resulting tachyonic instability. Namely, one can see that the sharpest-turn trajectory, $Ex5$ {\small ({\it solid line})}, has the greatest values of $|\Omega|_{t=t_{peak}}$ and $-m_s^2|_{t=t_{peak}}$\,. By comparison, $Ex7$ {\small ({\it dotted line})}, has an almost vanishing peak of $-m_s^2/H^2$\,. Note also that the position of the peak, $t_{peak}$\,, is clearly correlated with the position, $t_{max}$\,, of the local maximum of $\rho (t)$\,. On the other hand, the position of the local minimum of $\rho (t)$ is not related to any actual turn of the trajectory and thus is not affecting the shape of the function $\Omega (t)$\,. One can also notice that $t_{peak}$ is numerically closer to $t_{max}$ for the examples with higher peaks, than for those with lower peaks.

\begin{figure}[t]
\begin{center}
\hspace*{-0.2cm}
\includegraphics[scale=0.35]{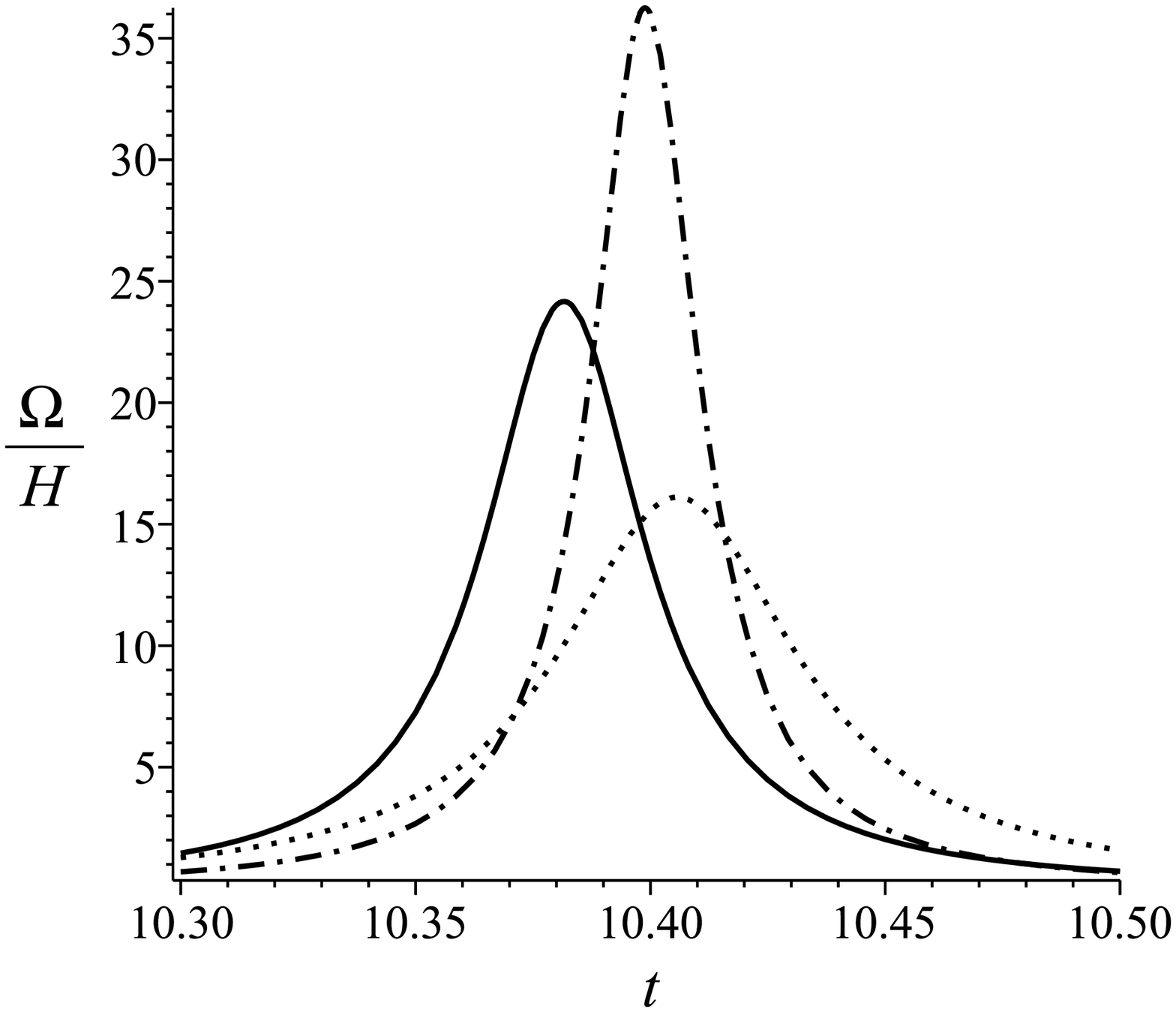}
\hspace*{0.4cm}
\includegraphics[scale=0.38]{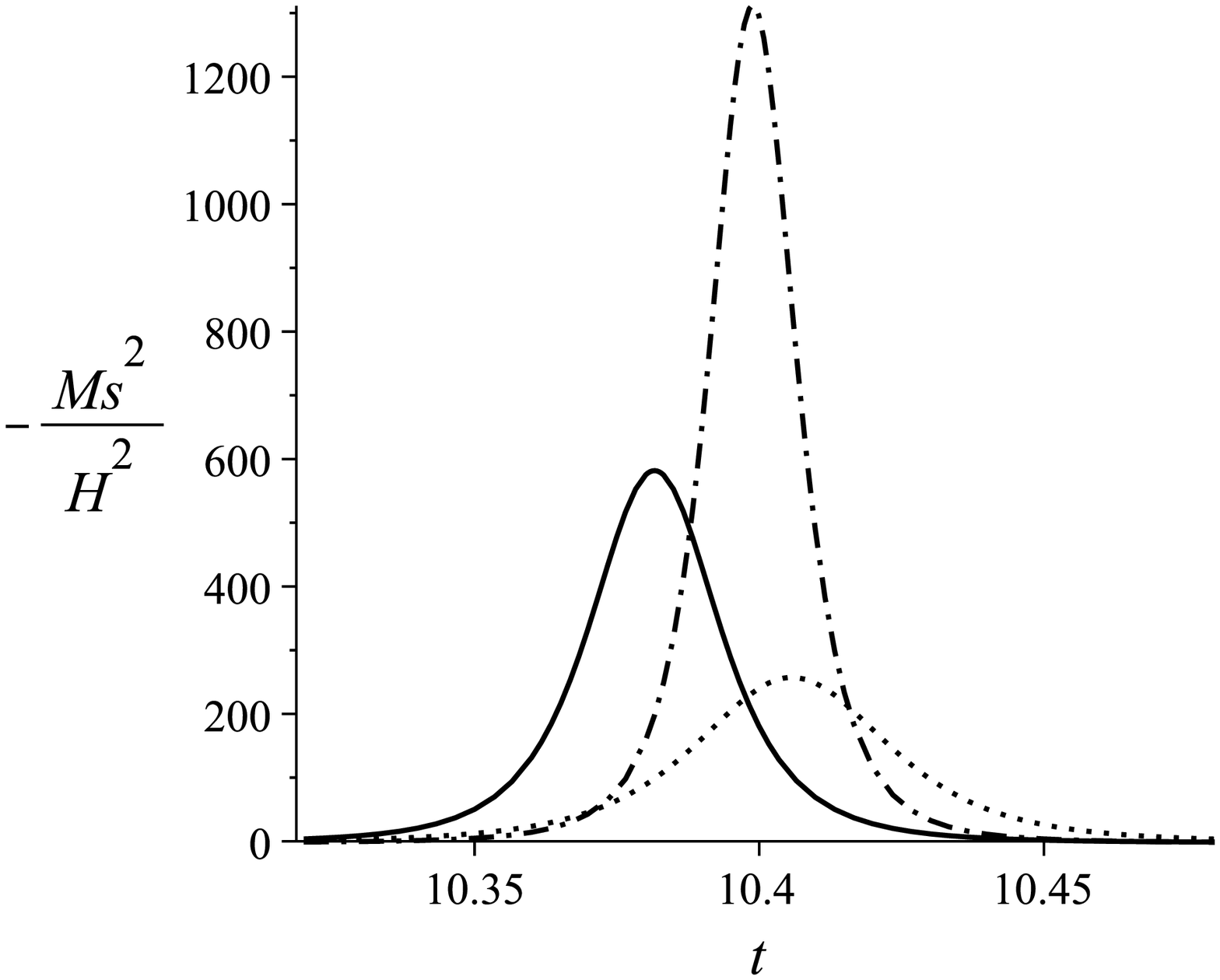}
\end{center}
\vspace{-0.7cm}
\caption{{\small The functions $\Omega/H$ and $-m_s^2/H^2$ for examples $Ex11$ ({\it dash-dotted line}), $Ex12$ ({\it solid line}) and $Ex13$ ({\it dotted line}) in the text. In these examples the number of e-folds that have occurred by the time of the peak is $N(t_{peak}) = 21$\,. And again, although not distinguishable on the graphs, $m_s^2$ is small and positive before and after the period of tachyonic instability.}}
\label{Tp_10}
\vspace{0.1cm}
\end{figure}
So far, we have considered only examples with $t_{peak} < 1$\,, which was convenient for illustrative purposes. In view of the analytical estimate (\ref{t_peak_l}) with $\kappa > 1$\,, it should not be surprising that it is easier to obtain $t_{peak} < 1$ than $t_{peak} > 1$. However, clearly, there are (infinitely many) examples with $t_{peak} > 1$\,. We have illustrated two such sets of examples on Figures \ref{Tp_5} and \ref{Tp_10}, demonstrating in the process that one can vary the height of $\Omega(t)|_{t=t_{peak}}$\,, as well as the resulting $-m_s^2 (t)|_{t=t_{peak}}$\,, as desired. On Figure \ref{Tp_5} we have plotted $\Omega/H$ and $-m_s^2/H^2$ for the following choices of constants (examples $Ex8$, $Ex9$ and $Ex10$ respectively): $C^u_0=6$, $C^v_0=1$, $C^v_1=-\frac{1}{5}$ and $C^w_1=\frac{1}{2}$ together with $C^w_0=-2.46$ {\small ({\it dash-dotted line})}, $C_0^w = -2.47$ {\small ({\it solid line})} and $C_0^w = -2.48$ {\small ({\it dotted line})}\,; in all three cases we have taken the value of $C_1^u$\,, determined by the positive root of (\ref{Constr_s}). In this set of examples $t_{peak} \sim 5$ and the number of e-folds $N \!= \!\int\!H dt$ at the peak is $N(t_{peak}) \!= \!11$\,, unlike in all of the above examples with $t_{peak} < 1$ for which $N(t_{peak}) \sim 1$\,. To facilitate comparison with \cite{PSZ,FRPRW}, we have also plotted $N(t)$ on Figure \ref{N_5}; that this is a linear function should be clear from the fact that the $\varepsilon$-parameter is exceedingly small, as discussed in Section \ref{TREM}. Interestingly, by directly comparing Figures \ref{Tp_5} and \ref{N_5}, we can see that the sharp turn occurs within a single e-fold, as in \cite{PSZ}. 

On Figure \ref{Tp_10} we have illustrated three examples ($Ex11$, $Ex12$ and $Ex13$ respectively) with $t_{peak} \sim 10$\,, obtained for $C^u_0=6$, $C^v_1=-\frac{1}{5}$, $C^w_1=\frac{1}{2}$ together with $C_0^v = 2.02$ and $C^w_0=-5.03$ {\small ({\it dash-dotted line})} , $C_0^v = 2.02$ and $C_0^w = -5.02$ {\small ({\it solid line})}, $C_0^v = 2.03$ and $C_0^w = -5.03$ {\small ({\it dotted line})}\,; again, in all three cases, we have taken the value of $C_1^u$ determined by the positive root of (\ref{Constr_s}). In these cases, with $t_{peak} \sim 10$\,, one has $\rho (t_{peak}) = 4 \times 10^{-16}$ and $N(t_{peak}) = 21$\,, demonstrating in particular that the number of e-folds can be varied as desired. Note that numerically $H=2$ again, to a great degree of accuracy. So $N = 2t$ for the examples on Figure \ref{Tp_10} as well, implying again that the sharp feature is contained within a single e-fold, as in \cite{PSZ}. 

All of the examples here with $t_{peak} > 1$\,, as well as many more, can be obtained in the following way. Start by choosing values for $C_{0,1}^{v,w}$ such that $\frac{C_0^v}{C_1^v} \approx \frac{C_0^w}{C_1^w}$\,, while $\frac{C_0^v}{C_1^v} \neq \frac{C_0^w}{C_1^w}$\,. Then take $C_0^u$ such that $(C_0^u)^2 \ge (C_0^v)^2 + (C_0^w)^2$\,, to ensure a positive-definite scale factor as discussed in the beginning of this Appendix. And, finally, solve (\ref{Constr_s}) for $C_1^u$\,. This procedure gives $t_{peak} \approx \frac{1}{\kappa}-\frac{C_0^w}{C_1^w}$\,, as well as $\Delta \theta \equiv |\theta_{fin} - \theta_{in}| \approx \pi$\,. 

\begin{figure}[t]
\begin{center}
\hspace*{-0.2cm}
\includegraphics[scale=0.33]{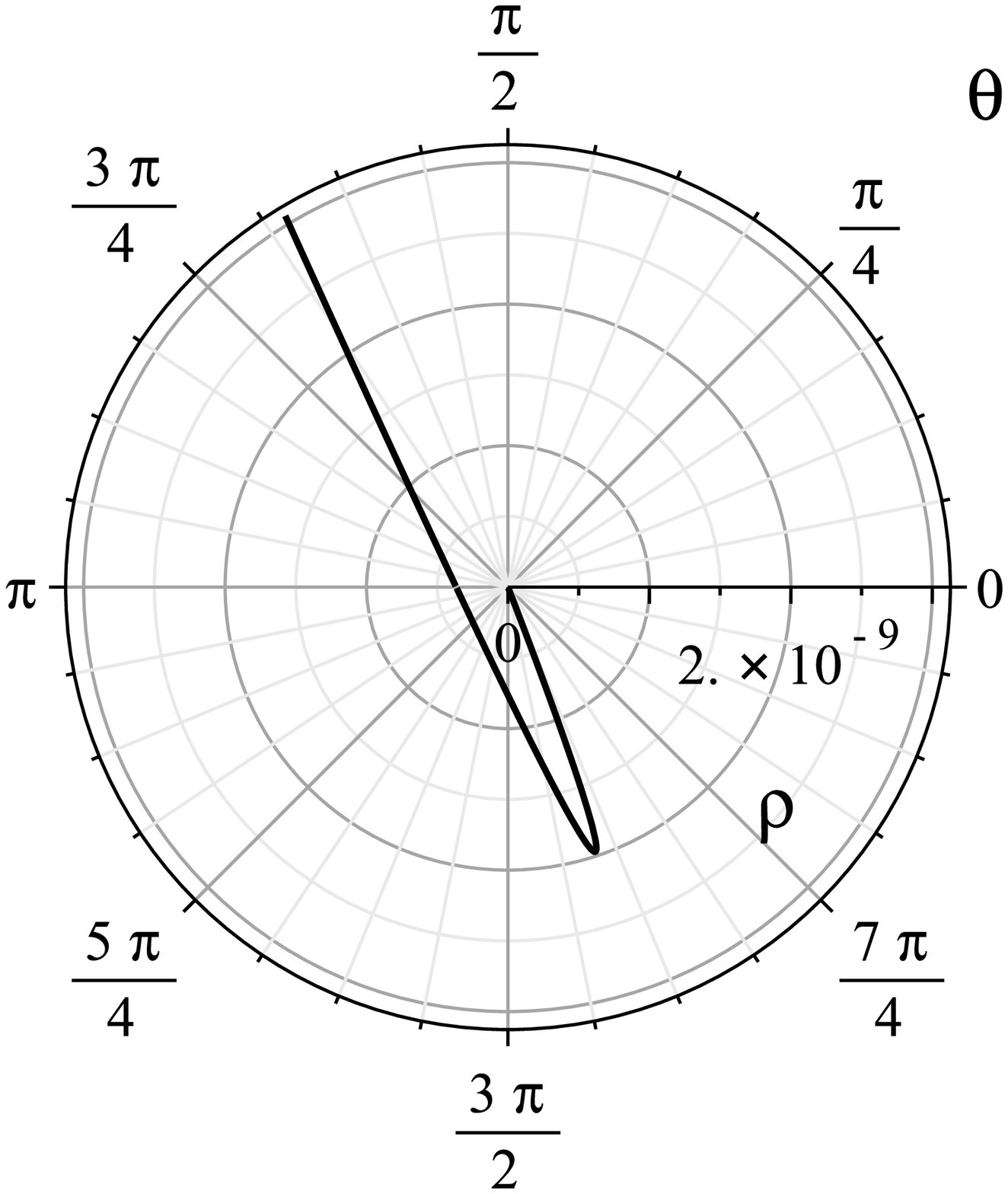}
\hspace*{0.4cm}
\includegraphics[scale=0.365]{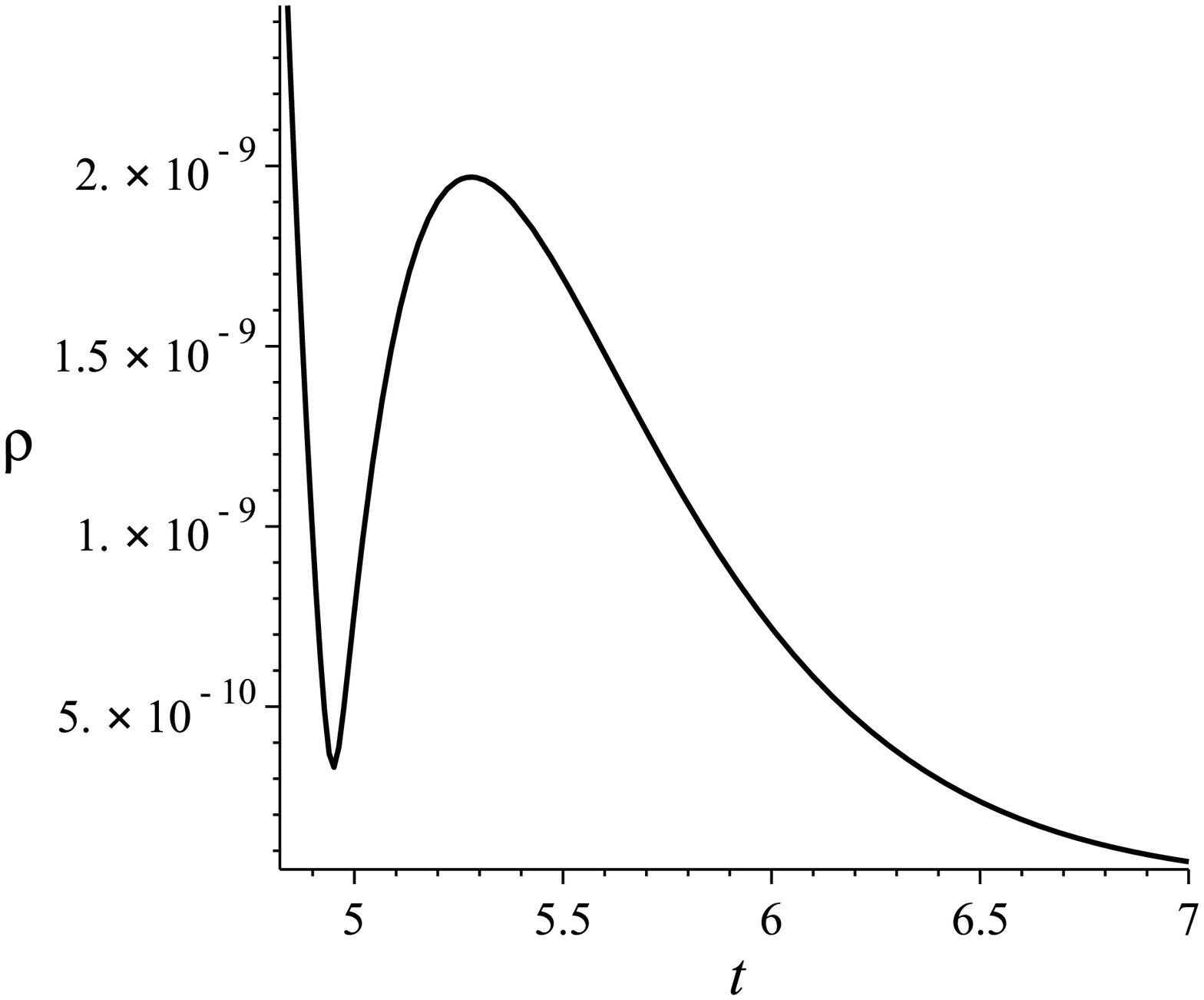}
\end{center}
\vspace{-0.7cm}
\caption{{\small The trajectory $\left( \rho (t) , \theta (t) \right)$ and the function $\rho (t)$ for the same example $Ex9$, as in the text and on Figure \ref{Tp_5}. In this example, $t_{peak} = 5.28$ and $\rho (t_{peak}) = 1.97 \times 10^{-9}$\,. The plot of the trajectory starts at $t = 4.82$\,, in order to make its shape visible.}}
\label{R_Th_l_5}
\vspace{0.1cm}
\end{figure}
Finally, on Figure \ref{R_Th_l_5} we have shown a typical example of a trajectory with $t_{peak} > 1$\,. The values of the constants in these plots are the same as in $Ex9$ above (the solid line on Figure \ref{Tp_5}). We have not started the plots at $t=0$\,, in order to be able to exhibit the features of the graphs. (Note that $\rho(t)|_{t=0} = 0.23$\,, which is orders of magnitude greater than $\rho(t)|_{t=t_{peak}}$.) Comparing Figures \ref{Tp_5} and \ref{R_Th_l_5}, one can notice again the correlation between $t_{peak}$ and $t_{max}$\,. It is important to make the following remark regarding the plot of the trajectory on Figure \ref{R_Th_l_5}, as well as any other trajectories with $C_0^w C_1^w < 0$\,. According to the $\theta (t)$ expression in (\ref{Sols_asc}), it would seem that $w=0$ is a singular point. However, this putative singularity is just a polar-coordinates issue and not a physical problem, as can be seen from (\ref{uvw_coord_tr}). Indeed, (\ref{Sols_asc}) was obtained by inverting (\ref{uvw_coord_tr}). Also, the function that enters all physics quantities is $\dot{\theta} (t) = \frac{v\dot{w}-\dot{v}w}{v^2+w^2}$\,, which is perfectly regular and even smooth.\footnote{Note that $v(t)$ and $w(t)$ in (\ref{Sols_uvw}) cannot vanish simultaneously at any moment of time, since by assumption the function $w(t)$ is different from $const \times v (t)$\,.} To circumvent the coordinate problem in (\ref{Sols_asc}), one can use Cartesian coordinates $x(t) = \frac{\rho v}{\sqrt{v^2 + w^2}}$ and $y(t) = \frac{\rho w}{\sqrt{v^2 + w^2}}$\,, in order to plot trajectories on the disk for examples with $C_0^w C_1^w < 0$ as on Figure \ref{R_Th_l_5}.

\end{document}